\documentclass[fleqn,usenatbib]{mnras}

\usepackage{newtxtext,newtxmath}

\usepackage{amsmath,bm}

\usepackage[T1]{fontenc}
\DeclareRobustCommand{\VAN}[3]{#2}
\let\VANthebibliography\thebibliography
\def\thebibliography{\DeclareRobustCommand{\VAN}[3]{##3}\VANthebibliography}

\DeclareUnicodeCharacter{2212}{-}

\usepackage{graphicx}	
\usepackage{bm}
\usepackage{amsmath}	

\usepackage{amssymb}	
\usepackage{multicol, blindtext} 
\usepackage{gensymb}
\usepackage{booktabs}

\usepackage{caption}
\DeclareCaptionFormat{cont}{#1 (cont.)#2#3\par}

\title[Tracking X-ray Outflows with Optical/IR Footprint Lines]{Tracking X-ray Outflows with Optical/IR Footprint Lines}

\author[Trindade Falcão et al.]{
Anna Trindade Falcão ,$^{1}$\thanks{E-mail: anna.trindade04@gmail.com}
S. B. Kraemer,$^{1}$
D. M. Crenshaw,$^{2}$
M. Melendez,$^{3}$
M. Revalski,$^{3}$
T. C. Fischer,$^{4}$
\newauthor
H. R. Schmitt,$^{5}$
T. J. Turner,$^{6}$
\\
$^{1}$Institute for Astrophysics and Computational Sciences, Department of Physics, The Catholic University of America, Washington, DC 20064, USA\\
$^{2}$Department of Physics and Astronomy, Georgia State University, Astronomy Offices, 25 Park Place, Suite 600, Atlanta, GA 30303, USA\\
$^{3}$Space Telescope Science Institute, 3700 San Martin Drive, Baltimore, MD 21218, USA\\
$^{4}$AURA for ESA, Space Telescope Science Institute, 3700 San Martin Drive, Baltimore, MD 21218, USA\\
$^{5}$Naval Research Laboratory, Washington, DC 20375, USA\\
$^{6}$Eureka Scientific, Inc., 2452 Delmer Street, Suite 100, Oakland, CA 94602-3017, USA\\
}

\date{Accepted XXX. Received YYY; in original form ZZZ}

\pubyear{2021}

\begin{document}
\raggedbottom
\label{firstpage}
\pagerange{\pageref{firstpage}--\pageref{lastpage}}
\maketitle

\begin{abstract}
We use Cloudy photoionisation models to predict the flux profiles for optical/IR emission lines that trace the footprint of X-ray gas, such as [Fe~X] 6375\AA~and [Si~X] 1.43$\mu$m. These are a subset of coronal lines, from ions with ionisation potential $\geq$ that of O~VII, i.e., 138eV. The footprint lines are formed in gas over the same range in ionisation state as the H and He-like of O and Ne ions, which are also the source of X-ray emission lines. The footprint lines can be detected with optical and IR telescopes, such as the \textit{Hubble Space Telescope}/STIS and \textit{James Webb Space Telescope}/NIRSpec, and can potentially be used to measure the kinematics of the extended X-ray emission gas. As a test case, we use the footprints to quantify the properties of the X-ray outflow in the Seyfert 1 galaxy NGC 4151. To confirm the accuracy of our method, we compare our model predictions to the measured flux from archival STIS spectra and previous ground-based studies, and the results are in good agreement. We also use our X-ray footprint method to predict the mass profile for the X-ray emission-line gas in NGC 4151 and derive a total spatially-integrated X-ray mass of $7.8(\pm 2.1) \times 10^{5}~M_{\odot}$, in comparison to $5.4(\pm 1.1) \times 10^{5}~M_{\odot}$ measured from a \textit{Chandra} X-ray analysis. Our results indicate that high-ionisation footprint emission lines in the optical and near-infrared can be used to accurately trace the kinematics and physical conditions of AGN ionised, X-ray emission line gas.

\end{abstract}



\section{Introduction}
\subsection{General Background}
\label{sec:introdution}
Active Galactic Nuclei (AGN) are powered by accretion of matter onto Supermassive Black Holes (SMBHs), with masses $\geq$ 10$^{6}$ ${\rm M\textsubscript{\(\odot\)}}$. This process of feeding the SMBH creates large amounts of electromagnetic radiation from the accretion disk of the black hole, which can accelerate winds, and may produce feedback \citep[e.g.,][]{begelman2004a}. These winds, if producing efficient feedback, evacuate the bulge of the host galaxy, quenching star formation, which is believed to produce the well-known relationship between the mass of the SMBH and the mass of the bulge \citep[e.g.,][]{gebhardt2000a}. \par 

In order to investigate how effective AGN feedback is on galactic-bulge scales, as required in a star-formation quenching, negative feedback scenario, it is important to quantify the mass outflows properties and their impact on the host galaxy. The physical properties of the gas, such as mass, mass outflow rates and kinetic energy can be estimated by photoionisation models \citep[e.g.,][]{crenshaw2007a}. These models, combined with emission-line studies, particularly those utilising high spatial-resolution \citep[e.g.,][]{fischer2017a, trindadefalcao2021a, revalski2021a}, provide the most accurate method to calculate critical quantities of these winds, such as the masses, velocities, outflow rates, and kinetic energies, as a function of distance from the SMBH. The strength of the AGN-driven winds can be quantified in the form of kinetic luminosity, $\dot E(r) = \frac{1}{2} \dot M_{out} v^2$, where the mass outflow $\dot M_{out}(r) = 4\pi rN_{H}\mu m_{p}C_{g}v_{r}$, and $r$ is the radial distance from the SMBH, $N_{H}$ is the column density, $\mu$ is the mean mass per proton, in this case =1.4, $m_{p}$ is the proton mass, $C_{g}$ is the global covering factor of the gas, and $v_{r}$ is the radial velocity. Studies for efficient feedback require $\dot E(r) \sim 0.5\% - 5\%$ of $L_{bol}$, the bolometric luminosity of the AGN radiating at near their Eddington limit \citep{dimatteo2005a, hopkins2010a}. In addition, the amount of kinetic energy deposited into the host galaxy rises rapidly with the velocity of the gas, since $\dot E \propto v^{3}$.\par

\begin{figure*}
  \centering
 \begin{minipage}[b]{0.45\textwidth}
  \includegraphics[width=8.65cm]{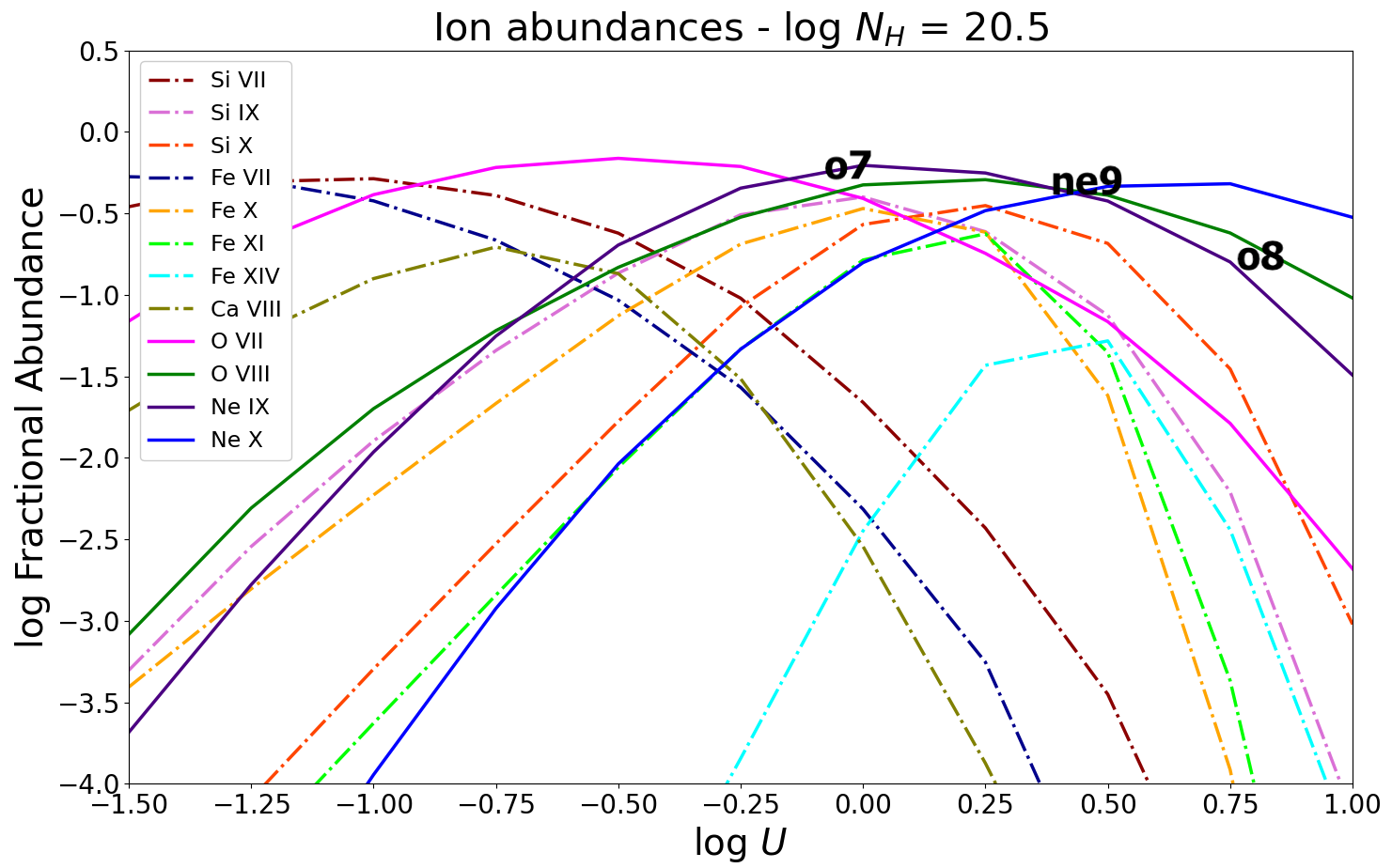}
 \end{minipage}\qquad 
 \begin{minipage}[b]{0.45\textwidth}
  \includegraphics[width=8.65cm]{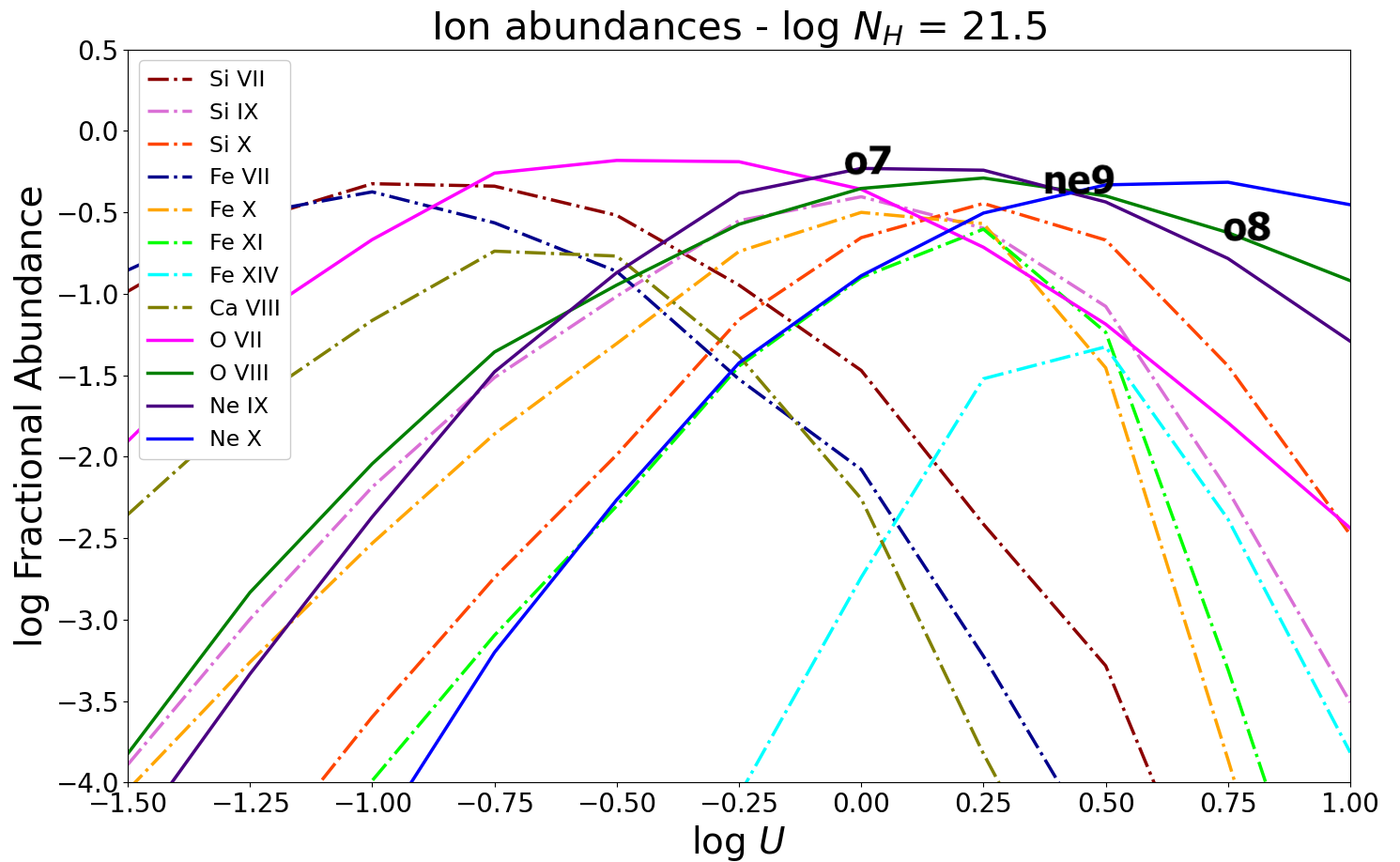}
 \end{minipage}\qquad 
 \begin{minipage}[b]{0.45\textwidth}
  \includegraphics[width=8.65cm]{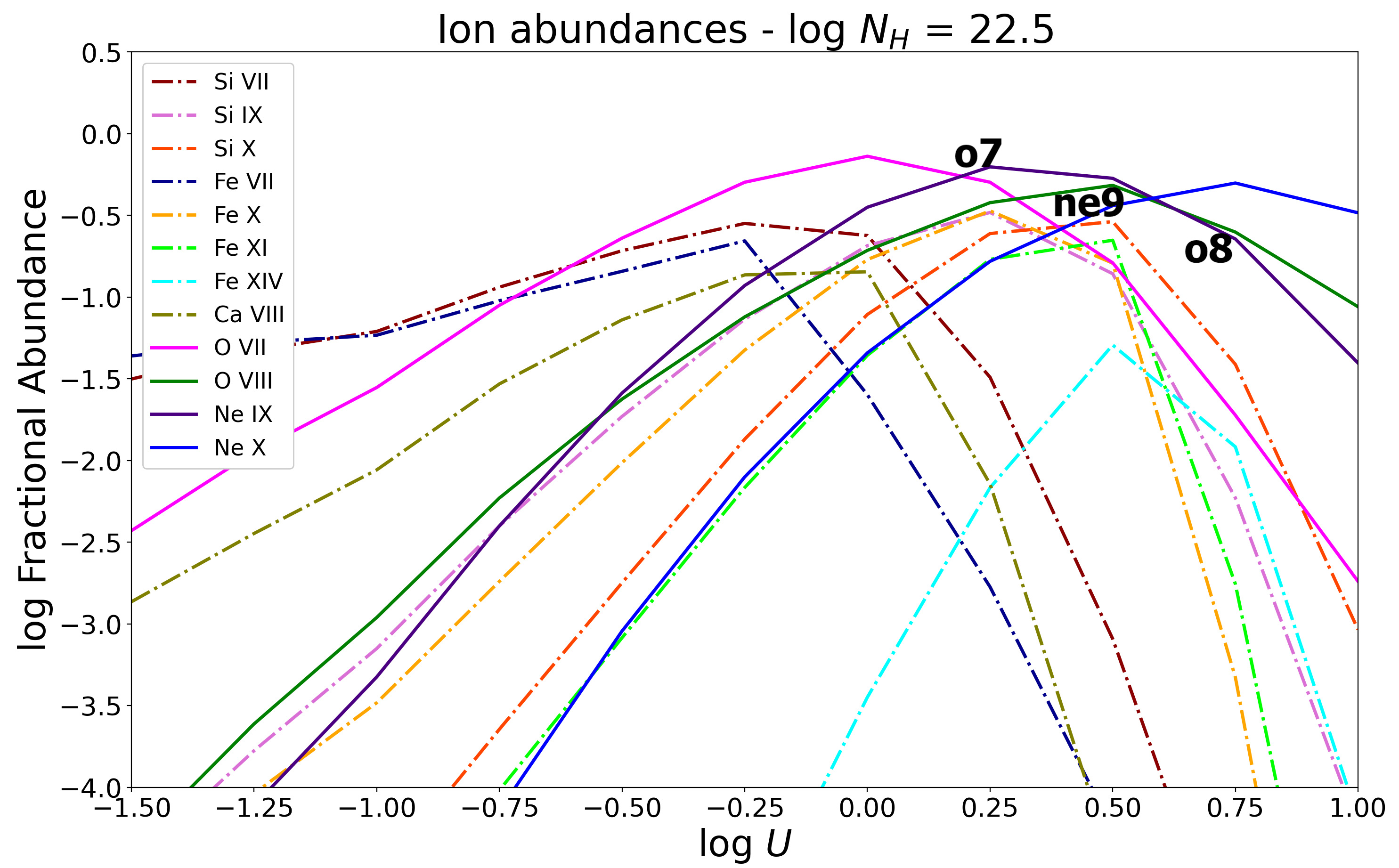}
\end{minipage}\qquad 
 \begin{minipage}[b]{0.45\textwidth}
  \includegraphics[width=8.65cm]{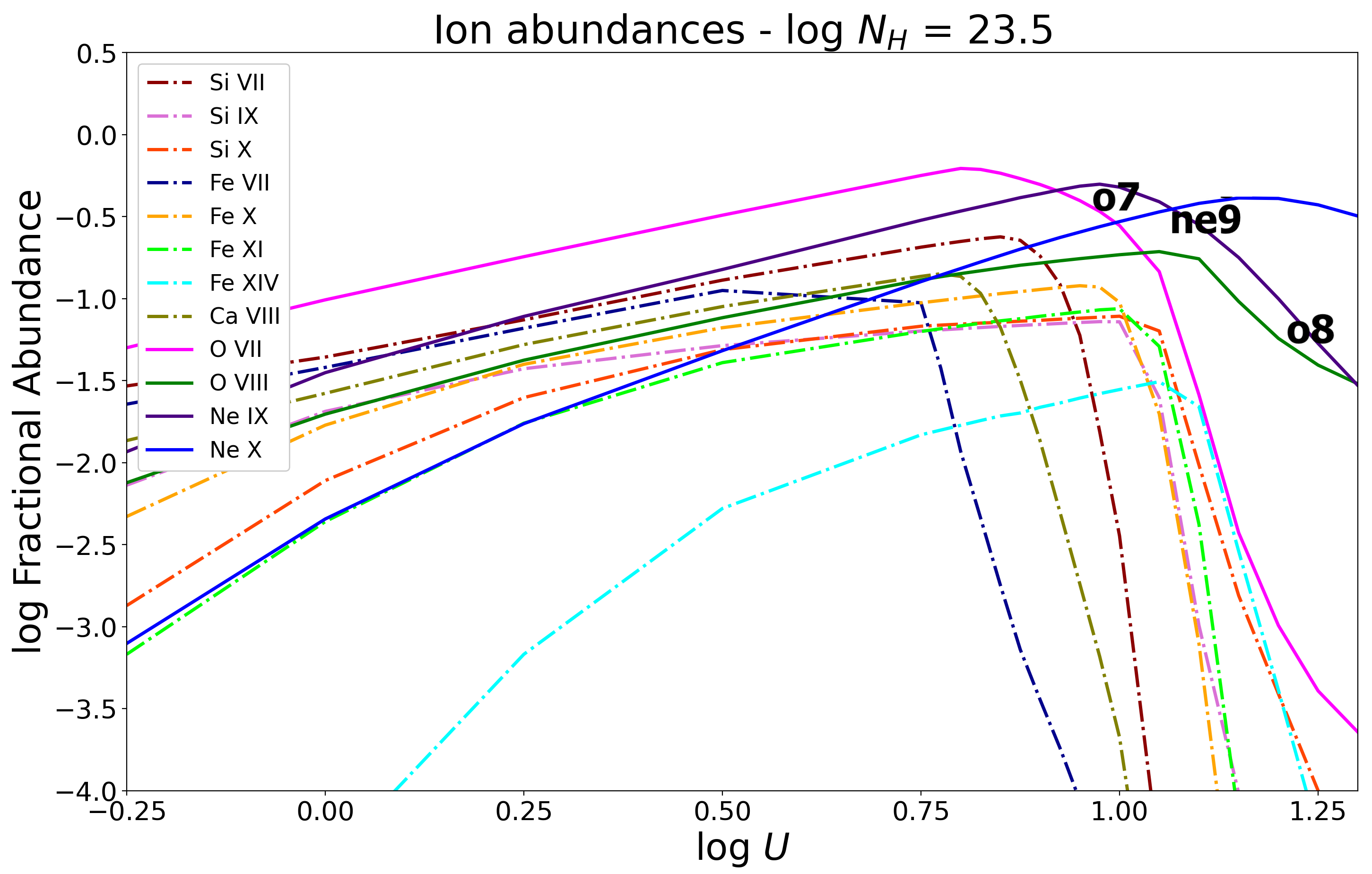}
 \end{minipage}\qquad 
\caption{Cloudy model predictions for the fractional abundances of relevant ionisation states of iron and silicon and H- and He-like oxygen and neon, for column densities of logN$_{H}$ = 20.5 (top-left), 21.5 (top-right), 22.5 (bottom-left), and 23.5 (bottom-right). Also, in boldface, we present the log$U$ and fractional abundances for which models predict the maximum fluxes for the X-ray emission-lines O~VII -f 22.1\AA, O~VIII Ly-$\alpha$, and Ne~IX -f 13.7\AA~ (as discussed in Section \ref{sec:study_footprints}). As shown here, optical emission lines from ions such as Fe~X, Fe~XI, Fe~XIV, and IR emission lines from ions such as Si~X will be formed in the X-ray emitting gas and will act as X-ray ``footprints".}
\label{fig:cloudy}
\end{figure*}

Our studies of mass outflow in nearby AGN show that, even though these outflows are very massive and inject considerable amounts of kinetic energy in the bulge of the host galaxy, they do not extend far enough to clear the bulge of gas \citep{fischer2018a} and also lack the power to do so \citep{trindadefalcao2021a}, since their $\dot E(r)/L_{bol}$ ratio does not reach the required 0.5\% for efficient feedback. Therefore, these results suggest that winds of optical emission-line gas are not an efficient form of AGN feedback \citep[e.g.,][]{trindadefalcao2021a}). However, these results do not take into account the role of higher-ionisation gas in this process of AGN feedback, which means that it is possible that X-ray winds are responsible for the dynamic effects we observe in nearby AGN \citep{trindadefalcao2021b}.\par 

Extended soft X-ray emission, co-located with the [O~III] emission-line gas, was detected by \textit{Chandra}/Advanced CCD Imaging Spectrometer (ACIS) in nearby Seyfert galaxies \citep{ogle2000a, young2001a}. In addition, \textit{Chandra} imaging has been used to map the Narrow Line Region (NLR) X-ray emission in several Seyferts \citep[e.g.,][]{bianchi2010a, gonzales2010a, wang2011a, wang2011b, wang2011c, maksym2019a}. By isolating bands dominated by specific emission-lines, these authors were able to derive constrains on the structure of the X-ray emission-line regions. Notably, \citet{wang2011a} and \citet{maksym2019a} suggest that there is evidence for shocks, which indicates interaction of the X-ray with the ISM of the host galaxy.\par 

Even though it possible to obtain data with good spectral resolution with \textit{Chandra}/HETG \citep[e.g.,][]{kallman2014a, kraemer2020a}, as well as kinematics and detailed physical insights, it is challenging to obtain detailed spatially-resolved information from these data. For instance, in their study of NGC 4151, \citet{kraemer2020a} were not able to get any accurate kinematic profiles from the HETG data, even for such a nearby AGN. Therefore, we have no means to determine what role this high ionisation gas plays in the process of AGN feedback.\par

Alternatively, a model for X-ray gas, characterised by a log$U$\footnote{$U= \frac{Q}{4\pi n_{H}r^{2}c}$, where, $n_{H}$ is the hydrogen number density, $r$ is the distance to the ionising source, $c$ is the speed of light, and the ionising luminosity, $Q = \int_{\nu_{0}}^{\infty} L_{\nu} \,d\nu \ $, where $h\nu_{0}$ = 13.6eV.} $\approx$ 0.0 predicts strong optical and IR emission lines (see Section \ref{sec:study_footprints}), which can be considered as "footprints" of the X-ray wind \citep[e.g.,][]{porquet1999a}. Since these footprint lines could be detected using ground-based
near-infrared telescopes \citep[e.g.,][]{rodriguez-ardila2011a, lamperti2017a} or the \textit{James Webb Space Telescope} \citep[][]{satyapal2021a} and \textit{HST}/STIS, it provides an
opportunity to accurately constrain the kinematics of the X-ray wind.

\begin{figure}
  \centering
  \includegraphics[width=8cm, height=5.3cm]{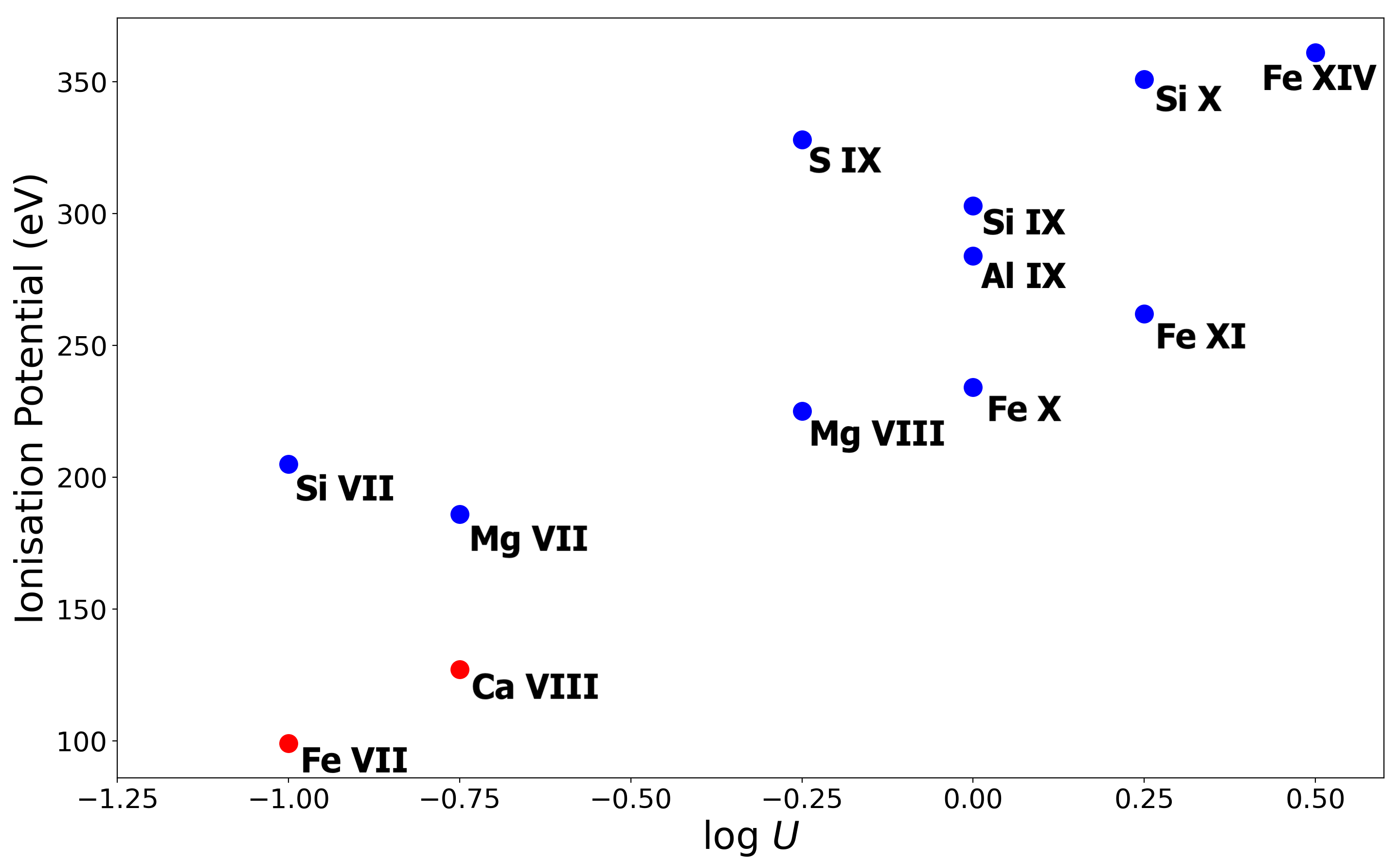}
\caption{Ionisation potential of the ions plotted against the ionisation parameter where the ion abundances peak. All the models are for logN$_{H}$=21.5.}
\label{fig:IP}
\end{figure}

\subsection{Evidence for X-ray feedback}
\label{x-ray_feedback}
In our most recent paper \citep[][]{trindadefalcao2021b}, we analysed the dynamics of mass outflows in the NLR of the QSO2 Mrk 34. We presented evidence for the presence of X-ray winds in the NLR of this QSO2 by performing an analysis based on the kinetic energy density of the [O~III]-emitting gas and the X-ray winds, and also by suggesting that high velocity gas is being accelerated in-situ via entrainment by X-ray winds \citep[for details see][]{trindadefalcao2021b}. \par 
We also used Cloudy photoionisation models \citep{ferland2017a} to calculate the integrated fluxes for some of the X-rays footprints, i.e., [Si~X] 1.43 $\mu$m and [Fe~X] 6375\AA. Even though our X-ray models provided predictions for the [Fe~X] that were consistent with the optical spectra, our parameterisation of the physical properties of the X-ray winds were determined indirectly, by its inferred dynamical effects and the distributions of the [O~III] gas in the NLR of Mrk 34. Therefore, although we were able to obtain results that were consistent with the X-ray emission detected in the \textit{Chandra}/ACIS image, this does not present itself as a well-constrained model of the X-ray wind.\par

\citet{kraemer2020a} analysed \textit{Chandra}/HETG spectra of the X-ray emission-line gas in NGC 4151. The zero-th order data show extended H- and He-like O and Ne, up to a distance $r \sim 200$ pc from the nucleus. They determined that the total mass of the ionised gas is $\approx$ $5.4 (\pm 1.1)\times10^{5}~{\rm M\textsubscript{\(\odot\)}}$. In addition, by assuming the same kinematic profile as that for the [O~III] gas, derived from the analysis by \citet{kraemer2000b} of \textit{HST}/STIS spectra, the peak X-ray mass outflow rate was $\approx$ 1.8 ${\rm M\textsubscript{\(\odot\)}}$ $yr^{−1}$, at r $\sim$ 150 pc. The total mass and mass outflow rates are similar to those determined using [O~III] \citep[][]{crenshaw2015a}, which implies that the X-ray gas is a major outflow component. In addition, the fact that it does not appear to be a drop in mass outflow rates for distances greater than 100 pc may indicate that the X-ray component has a greater effect on the host galaxy than the optical/UV gas and that X-ray winds might be a more efficient mechanism for AGN feedback. \par

Nevertheless, even though the use of \textit{Chandra}/HETG to characterise the X-ray outflows in NGC 4151 is an improvement compared to the ACIS imaging data used in \citealt{trindadefalcao2021b} to analyse the X-ray winds in Mrk 34, this analysis can only be used to obtain spatially-resolved spectra for the most nearby AGN. Additionally, the use of \textit{Chandra}/HETG is still not enough to fully characterise the extended emission, since this method does not provide accurate spatially resolved kinematics. For this reason \citet{kraemer2020a} made the assumption that the X-ray gas follows the [O~III] kinematics and used the [O~III] kinematic profile to track the X-ray gas in NGC 4151. In the present study we address this problem by analysing the X-ray outflows in NGC 4151 using a different methodology.\par

In this paper, we model the X-ray footprint lines using Cloudy photoionisation models, which allows us to construct radial flux profiles for these lines. We then compare our predictions to the results of \citet{kraemer2020a} for NGC 4151. In section \ref{sec:models} we present our photoionisation modeling analysis. In section \ref{sec:footprints} we present a comparison between our modeling results and the results of \citet{kraemer2020a}. Finally, in sections \ref{sec:discussion_1} and \ref{sec:discussion} we discuss our results and present our conclusions, respectively.

\section{The Study of Optical/IR X-ray Footprint Lines}
\label{sec:study_footprints}
In order to explore the relation between the footprint lines and the X-ray lines we generate a grid of Cloudy models using Mrk 34's SED and abundances, as discussed in \citealt{trindadefalcao2021a}. Our models consider a range in N$_{H}$ of 10$^{20.5}$ - 10$^{23.5}$, a range in log$U$ of (-1.50) - (1.00), and constant density, as the fractional abundance is not a strong function of density. Figure \ref{fig:cloudy} shows the Cloudy model predictions for the relative abundance of the Fe, Si, and Ca ions, compared to H and He-like O and Ne, as a function of the ionisation parameter and column density. The fact that these lines are formed in gas over the same range in ionisation state indicates
that they will be formed in gas that will also be the source of X-ray emission-lines, detected in the HETG or \textit{XMM-Newton}/Reflection Grating Spectrometer spectra. \par 

It is important to distinguish between coronal lines and footprint lines. Coronal lines are emission-lines formed from very highly ionised atoms, and they possess this name because they were first observed in the solar corona \citep[e.g.,][]{mazzalay2010a}. These lines are collisionally excited forbidden transitions within low-lying levels of highly ionised species with ionisation potentials (IP) $\geq$ 100 eV. They have been detected in the optical and IR spectra of all types of Seyfert galaxies \citep[e.g.,][]{seyfert1943a, nagao2000a, rodriguez-ardila2002a, landt2015a}, and represent one of the key gaseous components of the active nucleus. They appear to be approximately equally abundant in type 1 and type 2 AGN \citep[e.g.,][]{rodriguez-ardila2011a}.\par 

 On the other hand, we define footprint lines to be a subset of "coronal" lines from ions with IP $\geq$ 138 eV, i.e., that of O~VII, which peak over the same range in ionisation parameter as the H and He-like ions (see Figure \ref{fig:IP}). For instance, as shown in Figure \ref{fig:cloudy} (top right panel), the fractional abundance for Fe~VII, a ionisation state which produces a coronal line, peaks at log$U\approx$ -1.0. Even though its peak ionisation parameter is fairly close to that of O~VII, the peak flux of the OVII 22.1\AA~line is at significantly higher $U$. Additionally, the peak ionisation of O~VIII, Ne~IX, and Ne~X are all at
log$U$ $>$ 0. On the other hand, the fractional abundance for Fe~X, a ionisation state which produces a footprint line, peaks at log$U\approx$ 0.0, i.e., at the same ionisation parameter as the X-ray ions. It is also important to note here that the peak ionic abundance corresponds to the peak flux for collisionally excited lines, like the footprint lines, but not the X-ray lines. These are primarily due to recombination in photo-ionised plasma, as shown in Figure \ref{fig:cloudy}, hence peak at somewhat higher $U$ than where the ionic abundances is greatest.\par 

In addition, due to its high ionisation the gas is optically thin to the ionising radiation over a large range in column density, which means that our results are not a strong function of the column density of the gas. For example, in Figure \ref{fig:cloudy} we present the results for the Cloudy model predictions for the fractional abundances of relevant ionisation states of iron and silicon and H- and He-like oxygen and neon, for the following column densities: N$_{H}$ = 10$^{21.5}$ cm$^{-2}$ (top left panel), N$_{H}$ = 10$^{20.5}$ cm$^{-2}$ (top right panel), N$_{H}$ = 10$^{22.5}$ cm$^{-2}$ (bottom left panel), and N$_{H}$ = 10$^{23.5}$ cm$^{-2}$ (bottom right panel). As we can see, the ions from which the footprint lines originate and H- and He-like ions still coexist over the same range in ionisation parameter in a gas with different column densities, which means that our results do not change over a factor of $\gtrsim$ 1000 in N$_{H}$.\par 

Furthermore, we also analyse the effect of different SEDs on the fractional abundances of the relevant ionisation states shown in Figure \ref{fig:cloudy}. We use five different SEDs to perform this analysis, specifically, the SED presented by \citet{romano2004a} for the Narrow-Line Seyfert 1 Galaxy Arakelian 564, and the SEDs discussed by \citet{melendez2011a}. Since the choice of SED does not have a significant effect on the accuracy of our method, and to avoid interrupting the flow of the paper, we present these results in Appendix \ref{sec:appendix}.\par

An important point that should be addressed here is the location of [Si~VII] and [Mg~VII] in Figure \ref{fig:IP}. One could think that, since the ionisation parameter where their ionic abundances peak is $<$ log$U$=-0.5, these lines are not footprint lines, even though they satisfy our requirement for their IP, i.e., IP $\geq$ 138 eV. However, this phenomenon may be due to dielectronic recombination rates \citep[e.g.,][]{kraemer2004a, netzer2004a}. Specifically, if these rates are too low, the log$U$ values where the ionic abundances peak will also be too low. For example, a shift of 0.25 to 0.5 dex would be enough to place [Si~VII] and [Mg~VII] into the "footprint" range.

\begin{figure*}
  \centering
 \begin{minipage}[b]{0.4\textwidth}
  \includegraphics[width=7cm, height=4.3cm]{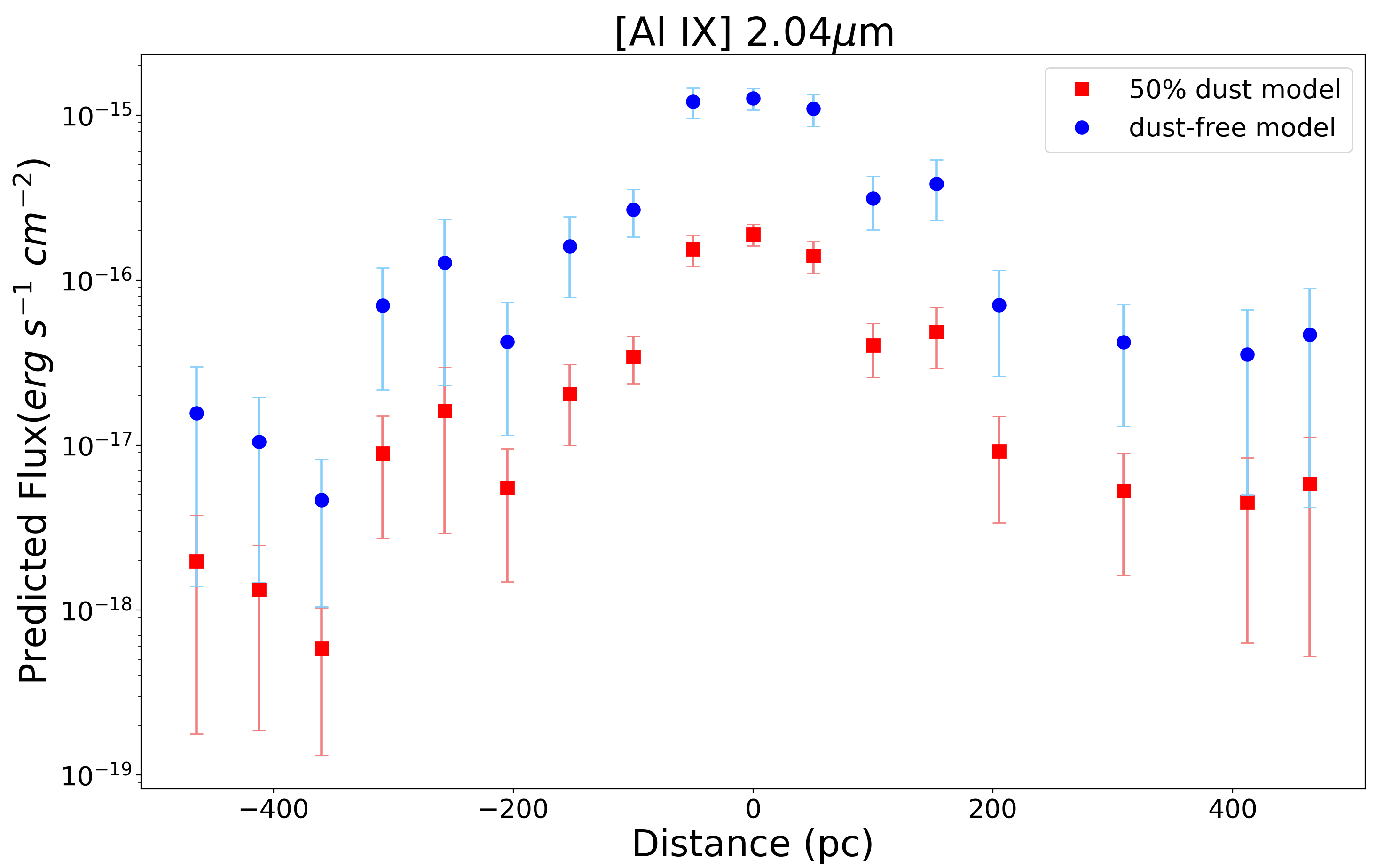}
 \end{minipage}\qquad 
 \begin{minipage}[b]{0.4\textwidth}
  \includegraphics[width=7cm, height=4.3cm]{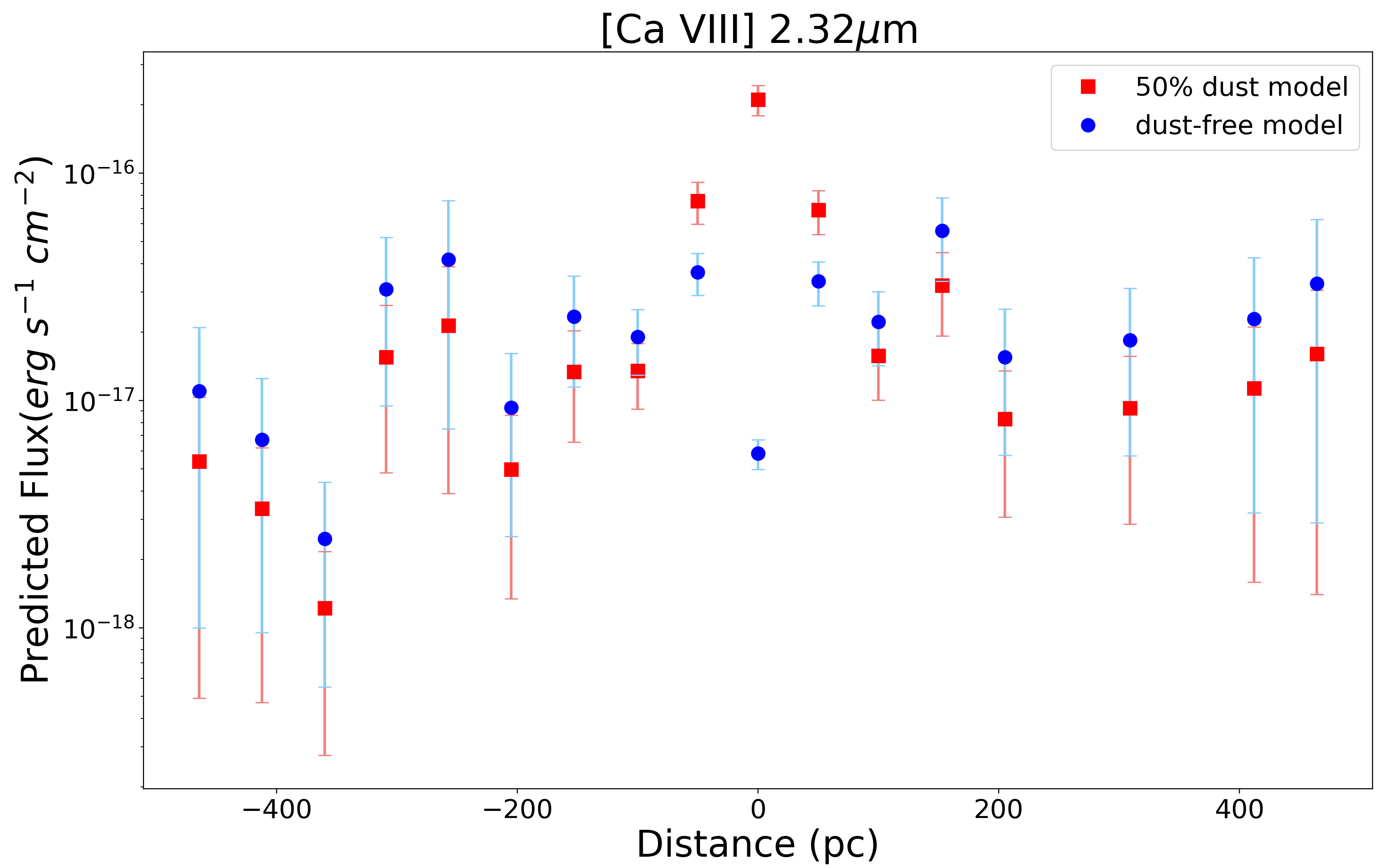}
 \end{minipage}\qquad 
 \begin{minipage}[b]{0.4\textwidth}
  \includegraphics[width=7cm, height=4.3cm]{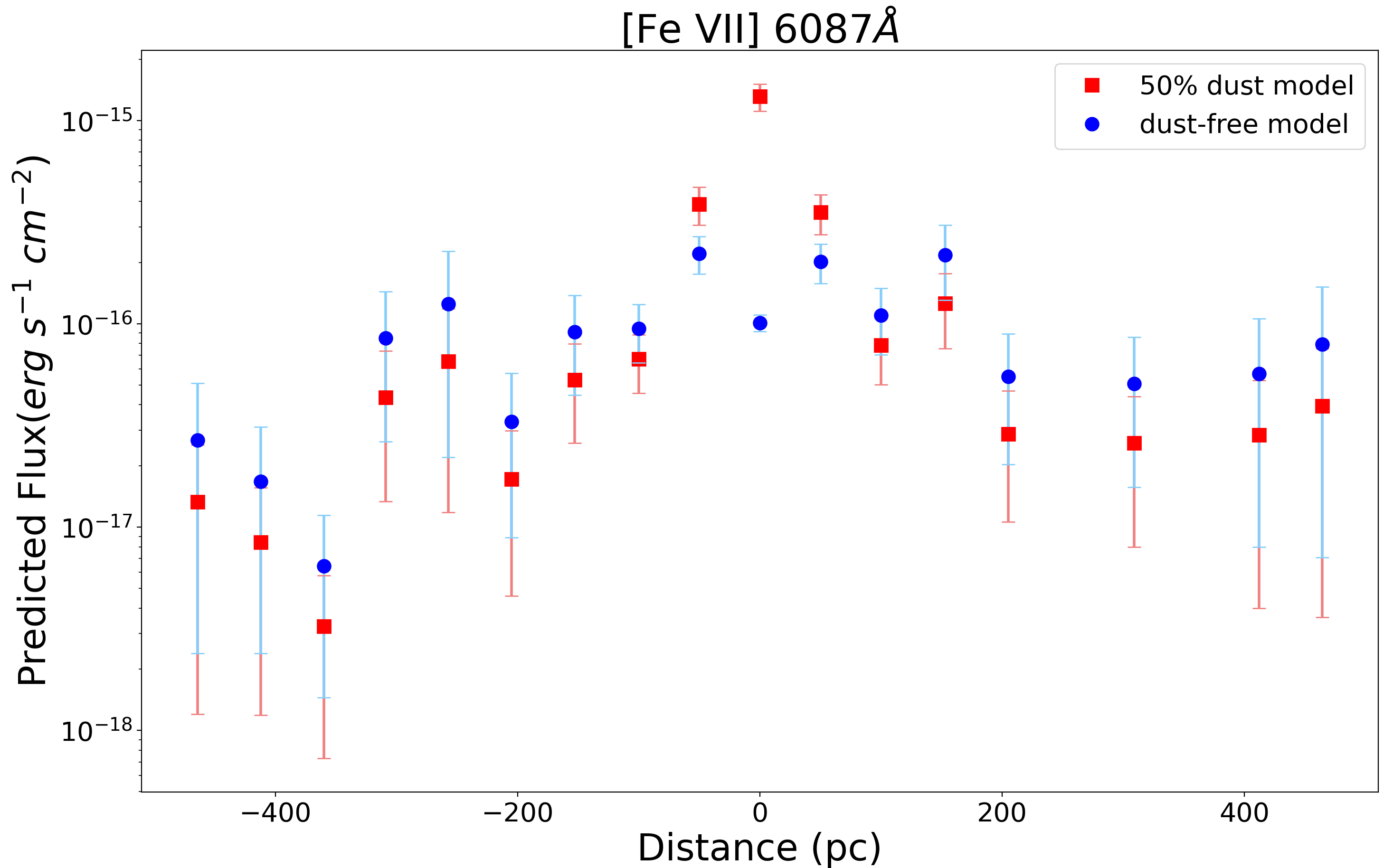}
\end{minipage}\qquad 
 \begin{minipage}[b]{0.4\textwidth}
  \includegraphics[width=7cm, height=4.3cm]{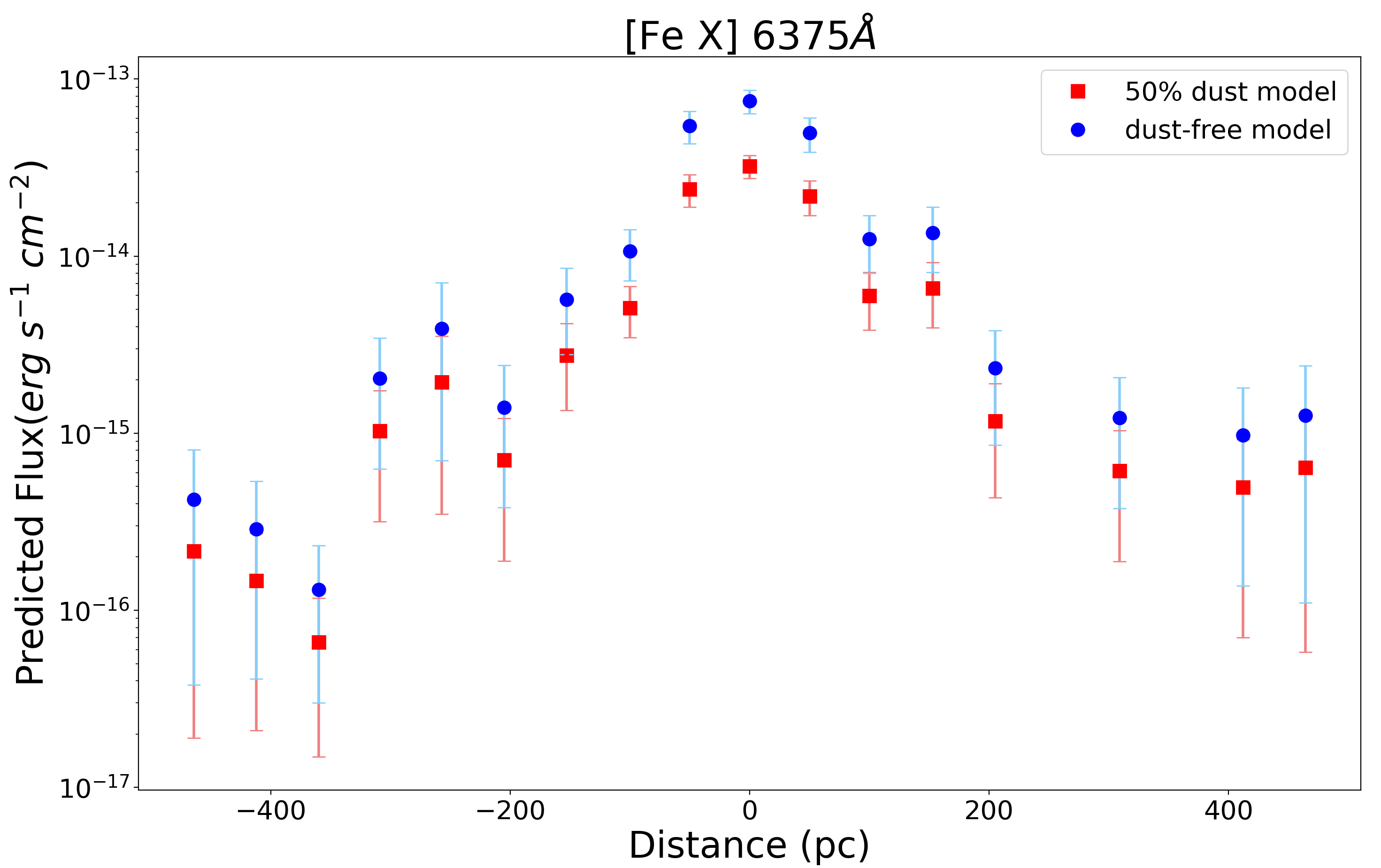}
 \end{minipage}\qquad 
\begin{minipage}[b]{0.4\textwidth}
  \includegraphics[width=7cm, height=4.3cm]{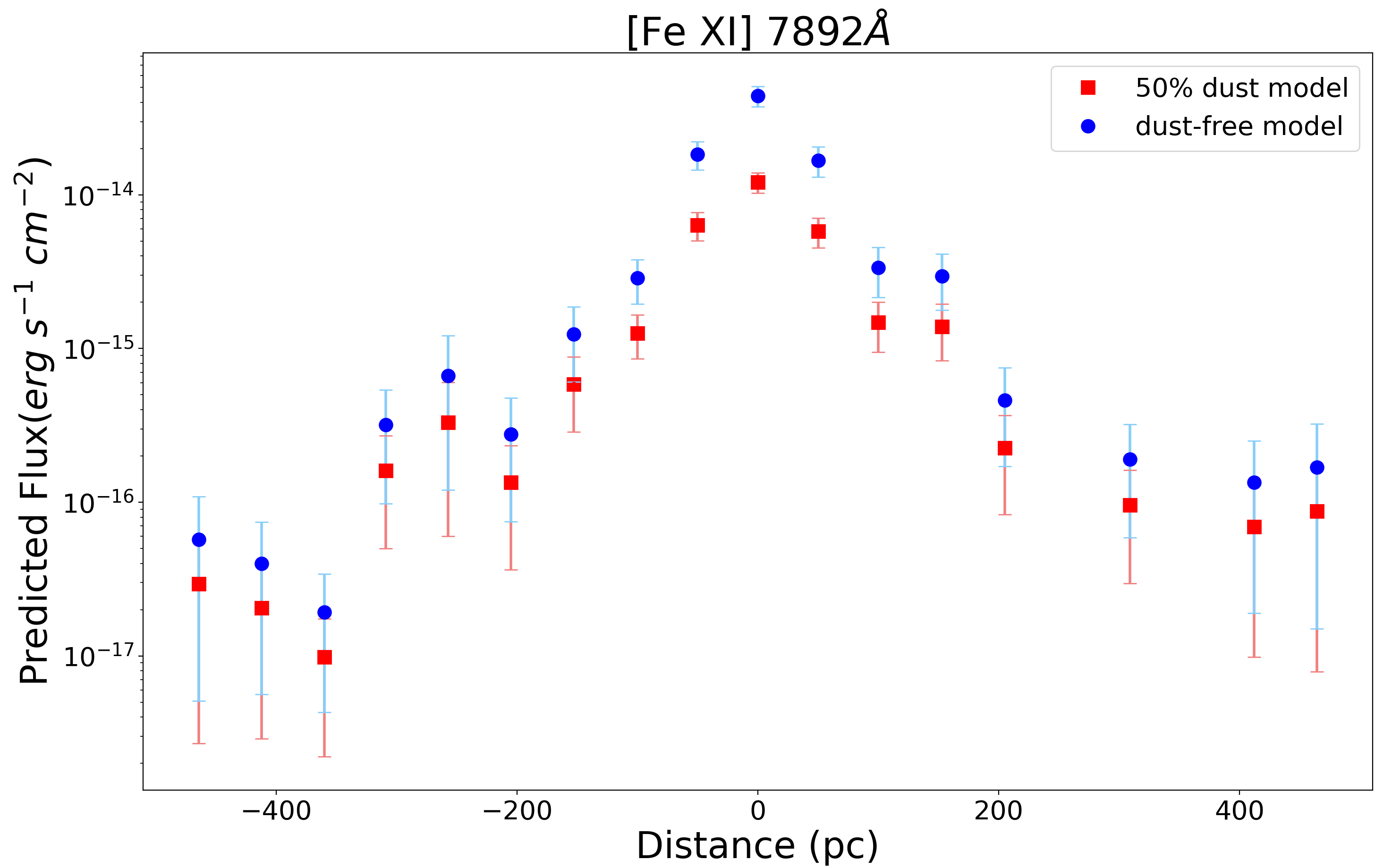}
\end{minipage}\qquad 
\begin{minipage}[b]{0.4\textwidth}
  \includegraphics[width=7cm, height=4.3cm]{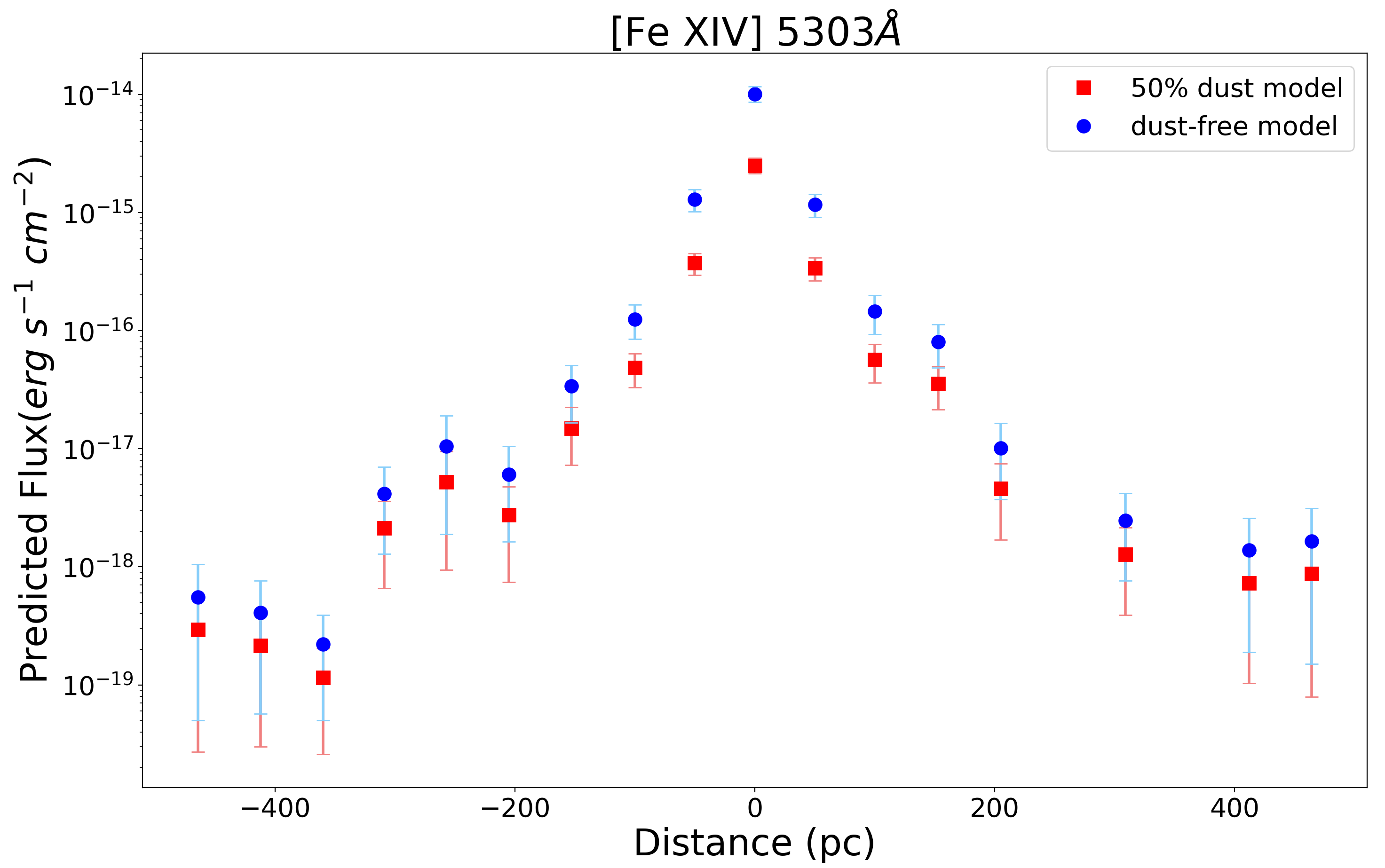}
\end{minipage}\qquad 
\begin{minipage}[b]{0.4\textwidth}
  \includegraphics[width=7cm, height=4.3cm]{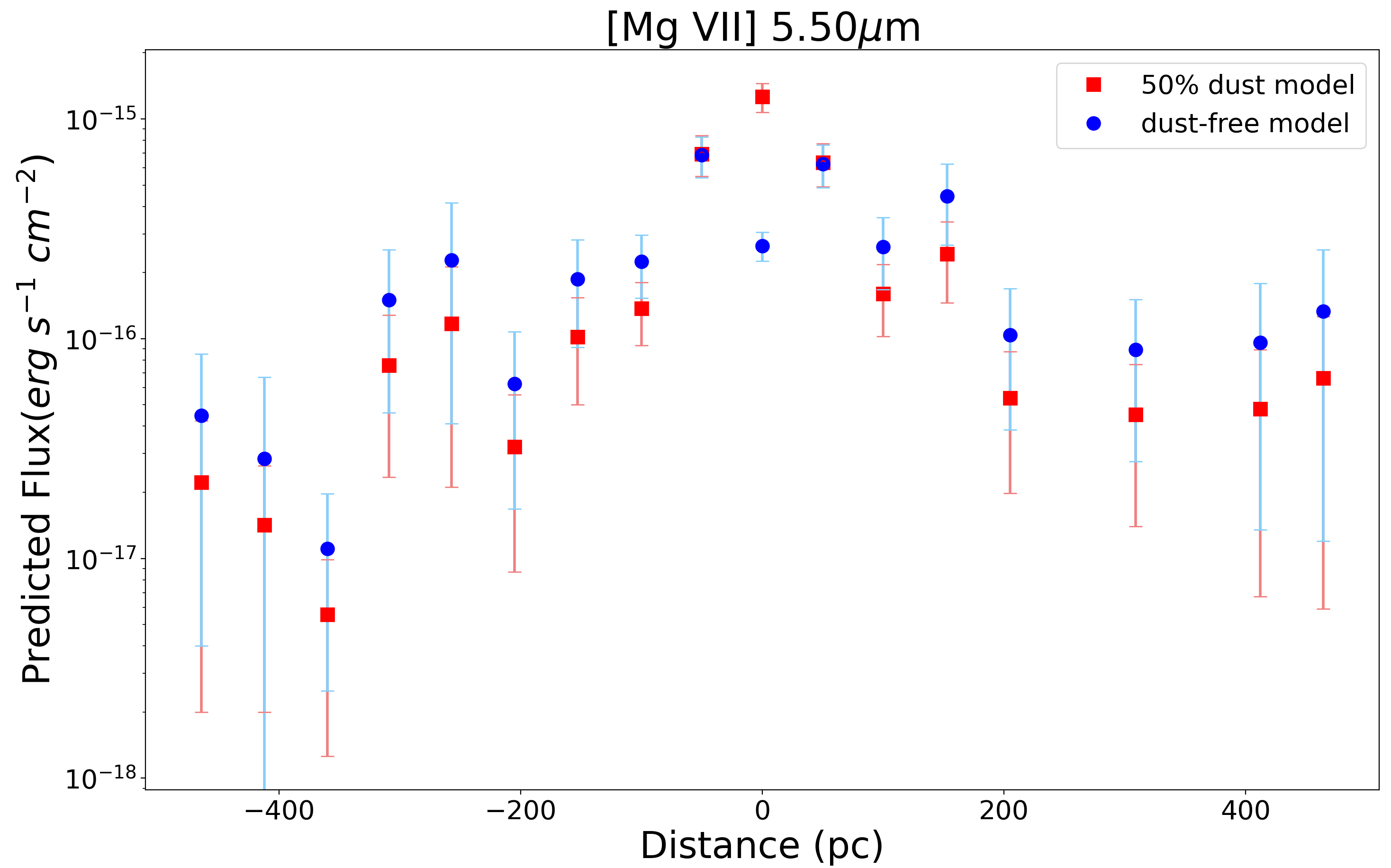}
\end{minipage}\qquad 
\begin{minipage}[b]{0.4\textwidth}
  \includegraphics[width=7cm, height=4.3cm]{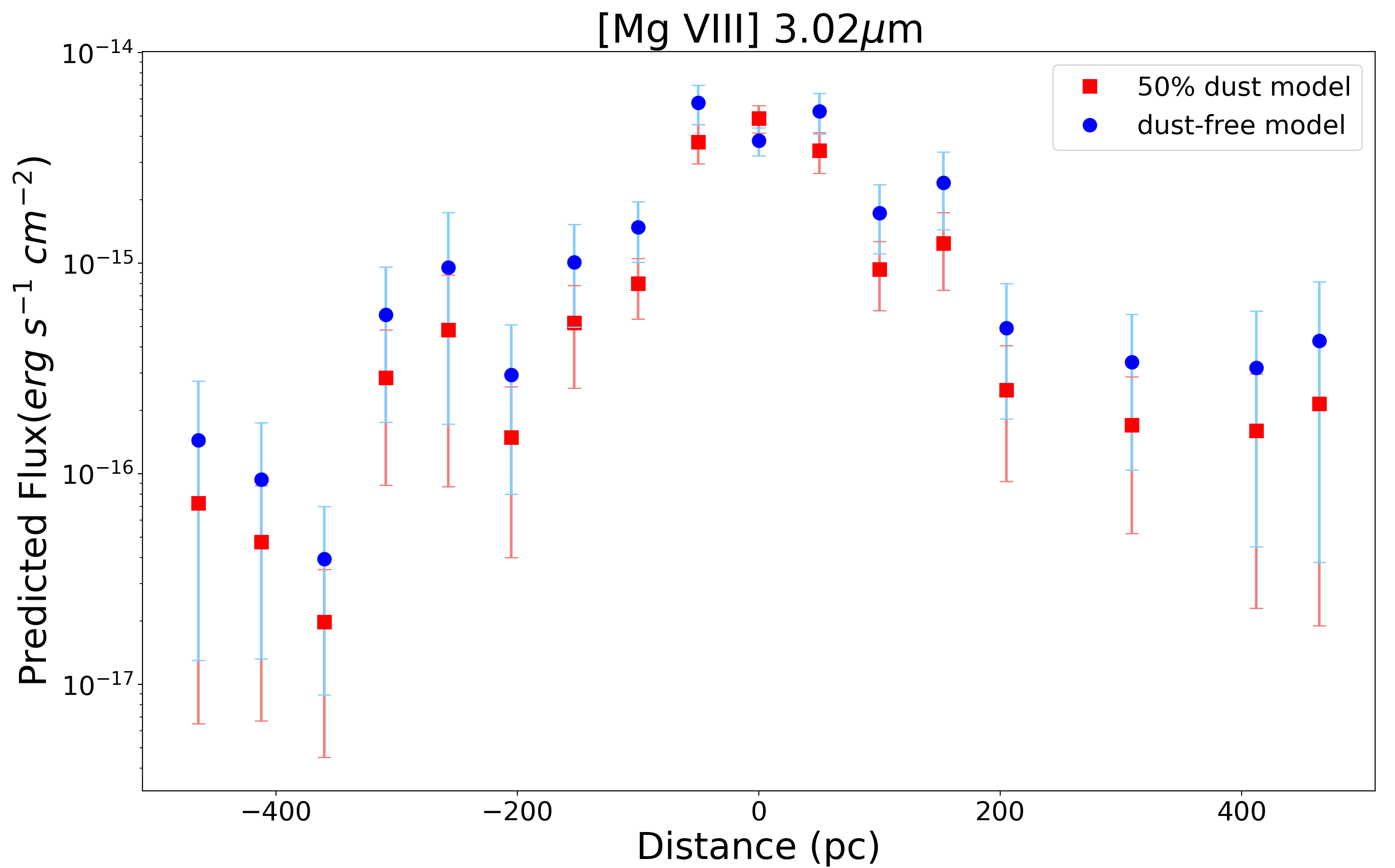}
\end{minipage}\qquad
\caption{Computed radial flux distributions for X-ray footprint lines [Al~IX] 2.04$\mu$m, [Ca~VIII] 2.32$\mu$m, [Fe~VII] 6078\AA, [Fe~X] 6375\AA, [Fe~XI] 7892\AA, [Fe~XIV] 5303\AA, [Mg~VII] 5.50$\mu$m, and [Mg~VIII] 3.02$\mu$m. The fluxes are obtained using the analysis described in section \ref{sec:models}. The dust-free models are plotted in red and the 0.5 ISM models are plotted in blue. The models assume $\delta r/r$ $<$ 1 or $\delta r$ fixed at 50pc, where $\delta r$ is the deprojected width for each extraction bin. The uncertainties were obtained based on the S/N of the fluxes in the 0th order image from \citealt{kraemer2020a}. The positive and negative distances correspond to point SW and NE of the nucleus, respectively.}
\label{fig:fluxes1}
\end{figure*}

\renewcommand{\thefigure}{\arabic{figure} (Cont.)}
\addtocounter{figure}{-1}

\begin{figure*}
\begin{minipage}[b]{0.4\textwidth}
  \includegraphics[width=7cm, height=4.3cm]{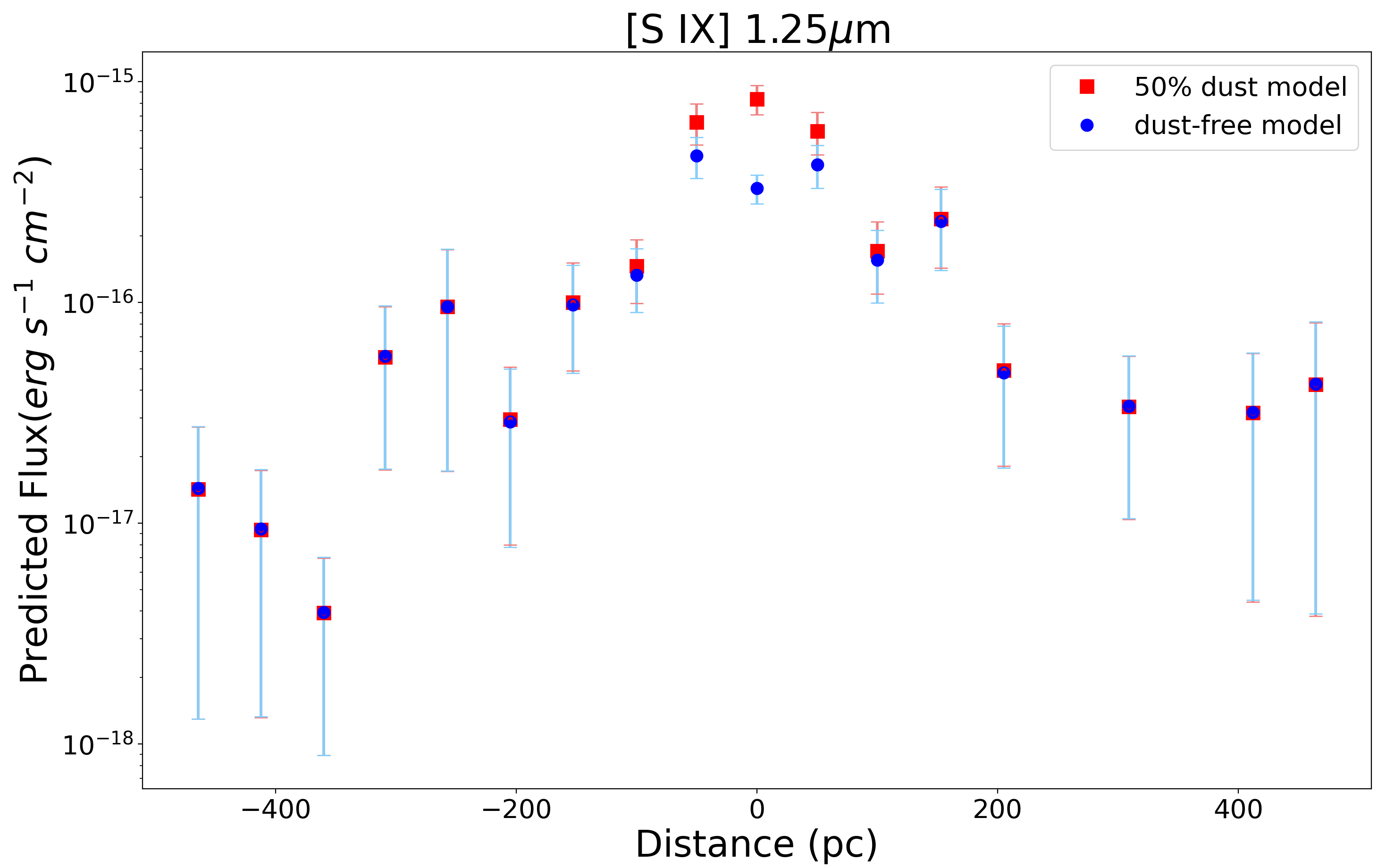}
\end{minipage}\qquad 
\begin{minipage}[b]{0.4\textwidth}
  \includegraphics[width=7cm, height=4.3cm]{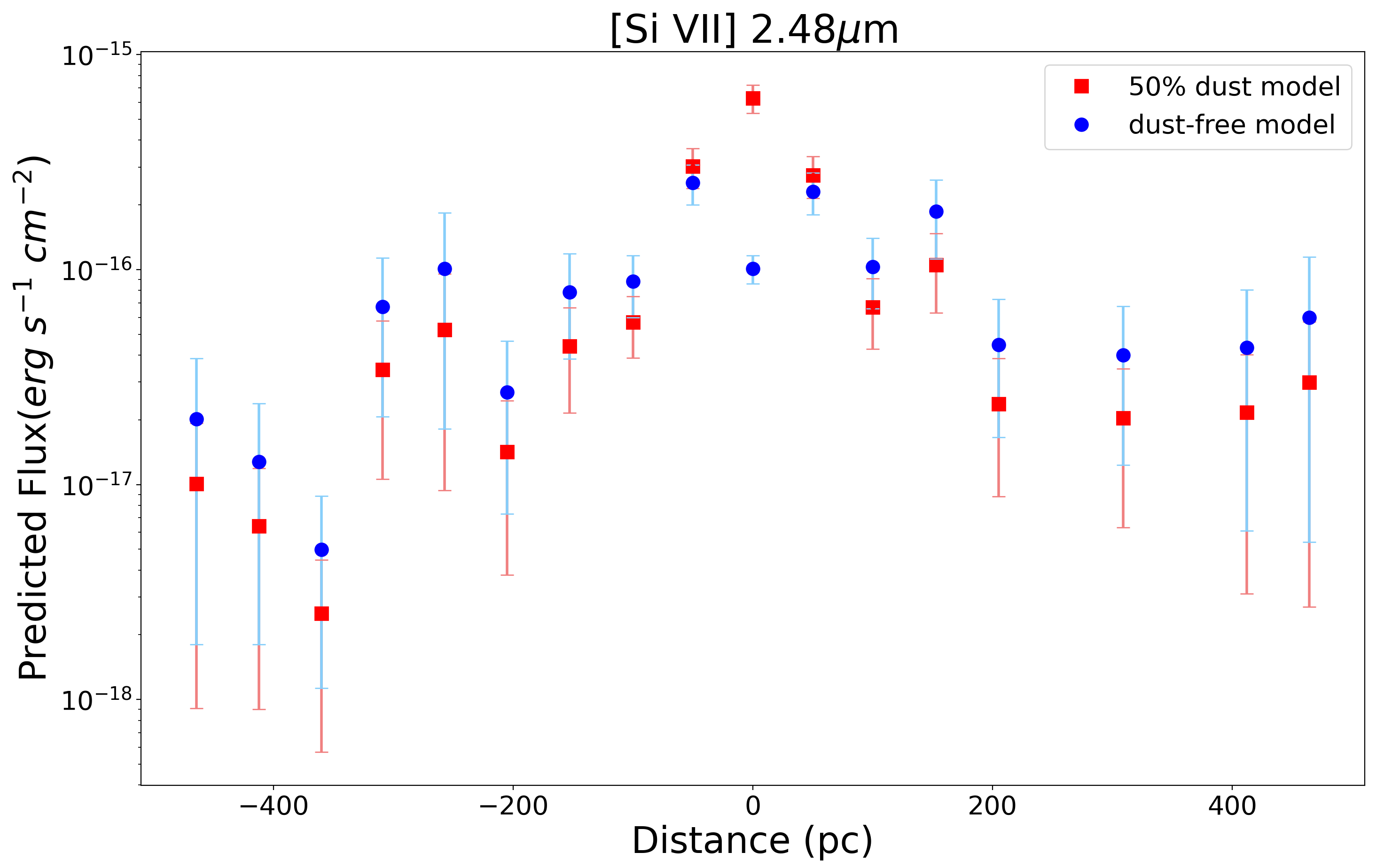}
\end{minipage}\qquad
\begin{minipage}[b]{0.4\textwidth}
  \includegraphics[width=7cm, height=4.3cm]{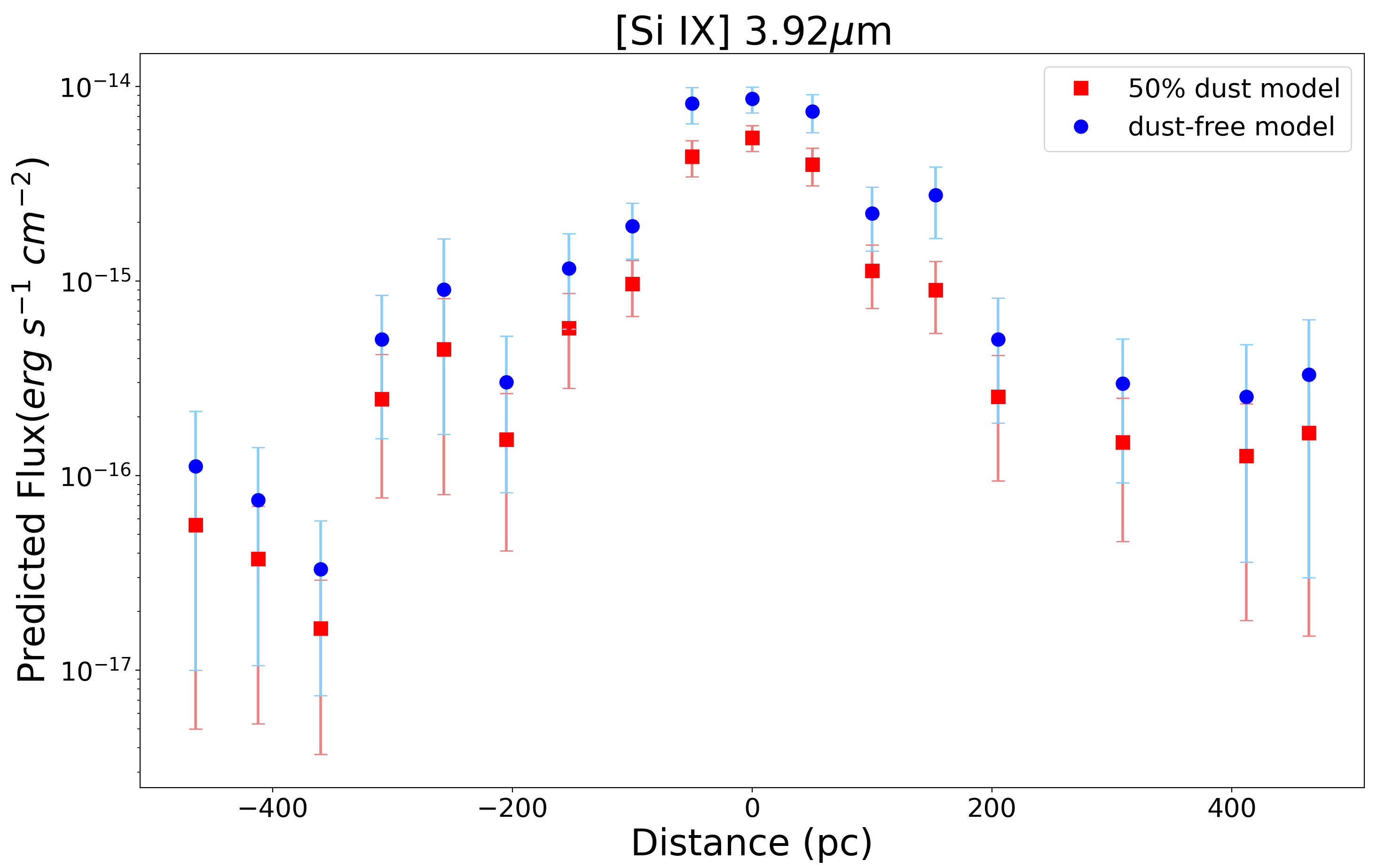}
\end{minipage}\qquad 
\begin{minipage}[b]{0.4\textwidth}
  \includegraphics[width=7cm, height=4.3cm]{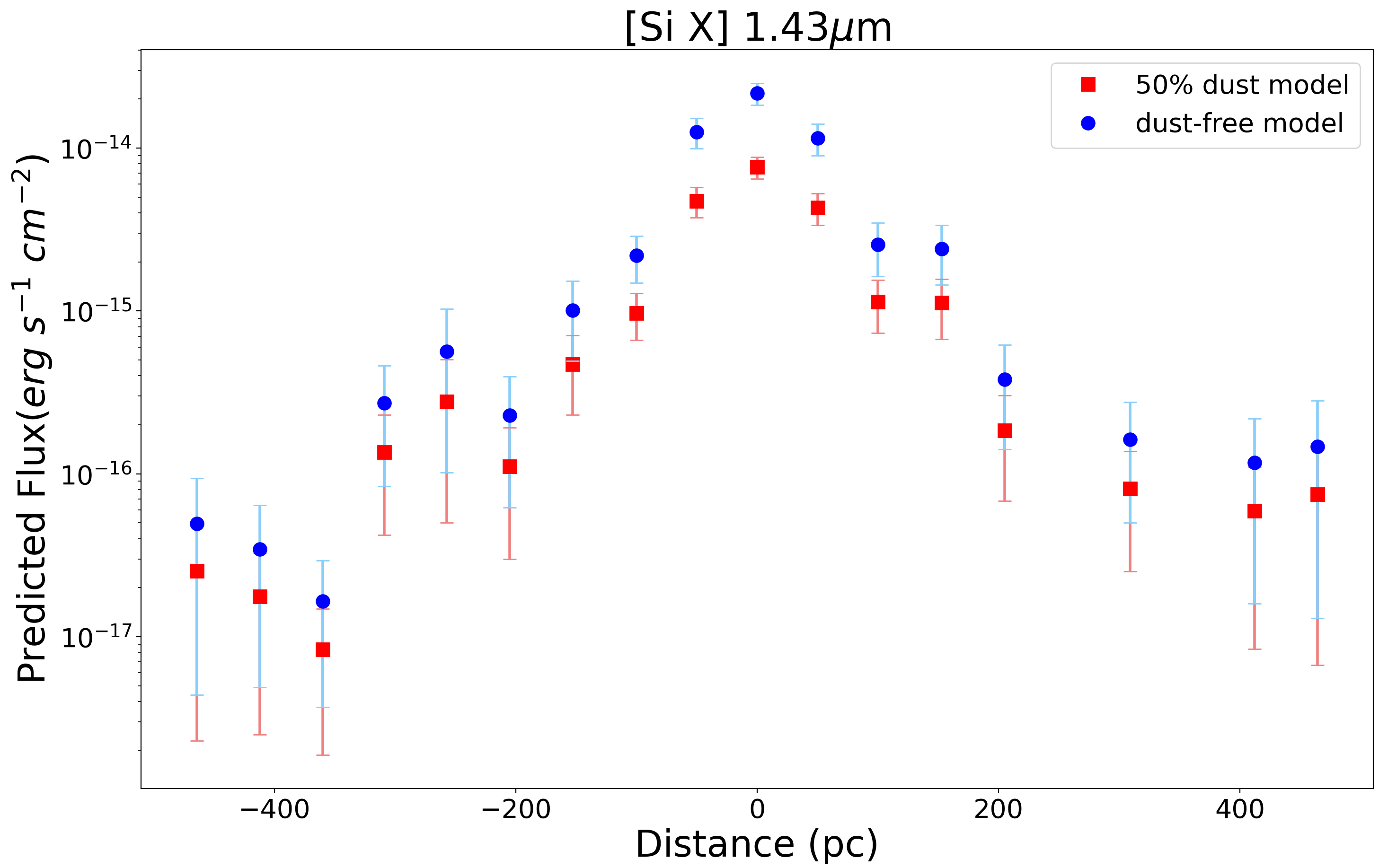}
\end{minipage}\qquad
\caption{Computed radial flux distributions for X-ray footprint lines [S~IX] 1.25$\mu$m, [Si~VII] 2.48$\mu$m, [Si~IX] 3.92$\mu$m, and [Si~X] 1.43$\mu$m.}

\label{fig:fluxes1_1}
\end{figure*}

\section{Photionisation Modeling Analysis}
\label{sec:models}

\subsection{Photoionisation Models}
\label{sec:photoionisation}
For our analysis, we generate photoionisation model grids\footnote{The model predictions for the ions fractional abundances present in Figure \ref{fig:cloudy} use the SED of Mrk 34 \citep{trindadefalcao2021a} and parameters as discussed in the caption of the figure.}, computed using Cloudy \citep[version 17.00;][]{ferland2017a}. We assume a continuum source with a SED that can be fitted using a power law of the form $L_{\nu} \propto \nu^{-\alpha}$. For our study we adopt the same SED used by \citet{kraemer2020a} in their study on NGC 4151, i.e. 
\medskip

 $ \alpha$ = 1.0 for $1\times 10^{-4}$ $\leq$ ${ h \nu}$ $\leq$ 13.6 eV;  \par
 $ \alpha$ = 1.3 for 13.6 eV $\leq$ ${ h \nu}$ $\leq$ 500 eV;  \par
 $ \alpha$ = 0.5 for 500 eV $\leq$ ${ h \nu}$ $\leq$ 30 keV; \par
 
 In addition, the choice of input parameters, specifically, the radial distances of the emission-line gas with respect to the central source ($r$), number density ($n_{H}$), the ionisation parameter ($U$), and column density ($N_{H}$) of the gas, is that to match the input parameter used by \citet{kraemer2020a}, which we will briefly describe. \par 

\renewcommand{\thefigure}{\arabic{figure}}
\begin{figure*}
  \centering
 \begin{minipage}[b]{0.4\textwidth}
  \includegraphics[width=7cm, height=4.3cm]{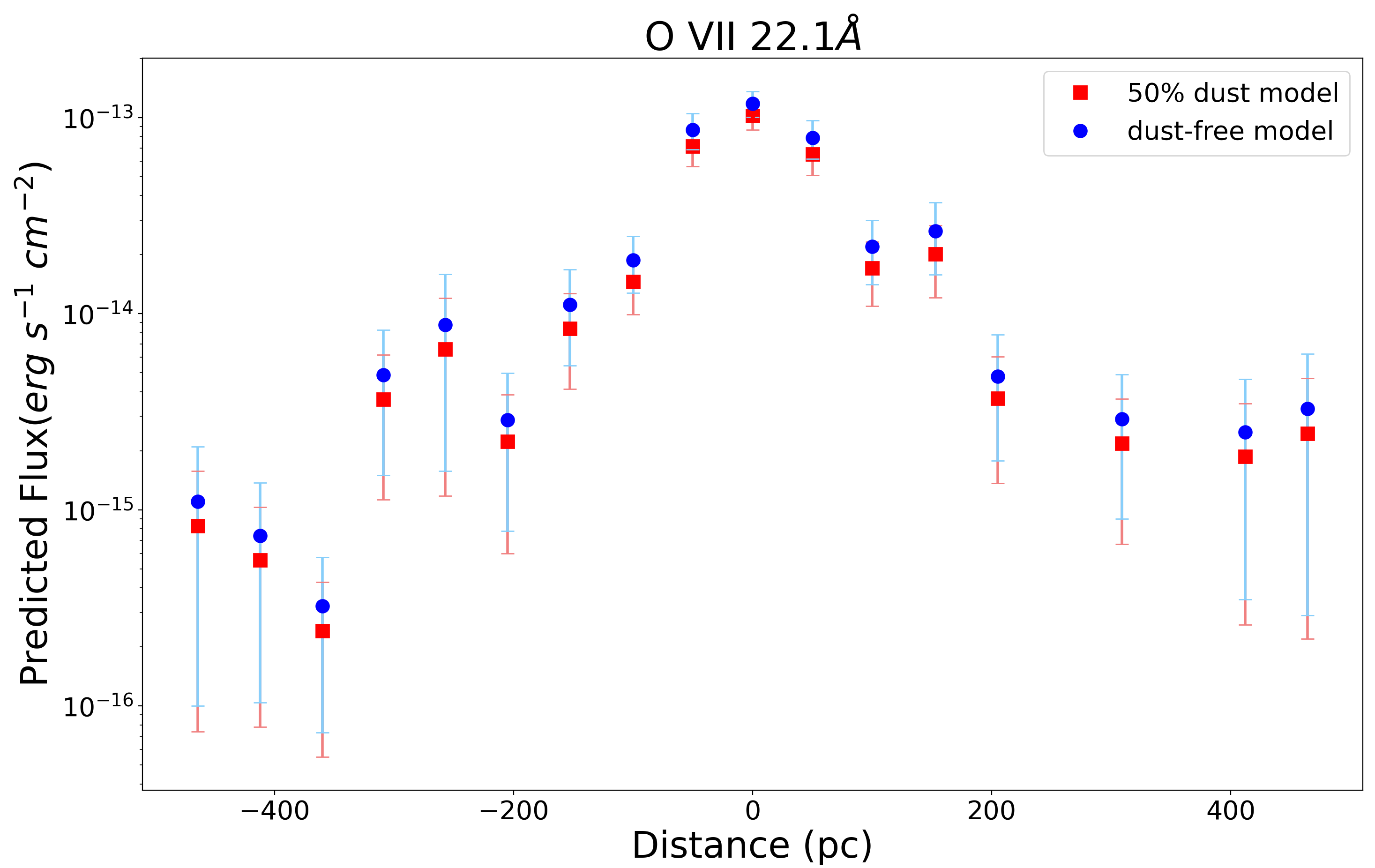}
 \end{minipage}\qquad 
 \begin{minipage}[b]{0.4\textwidth}
  \includegraphics[width=7cm, height=4.3cm]{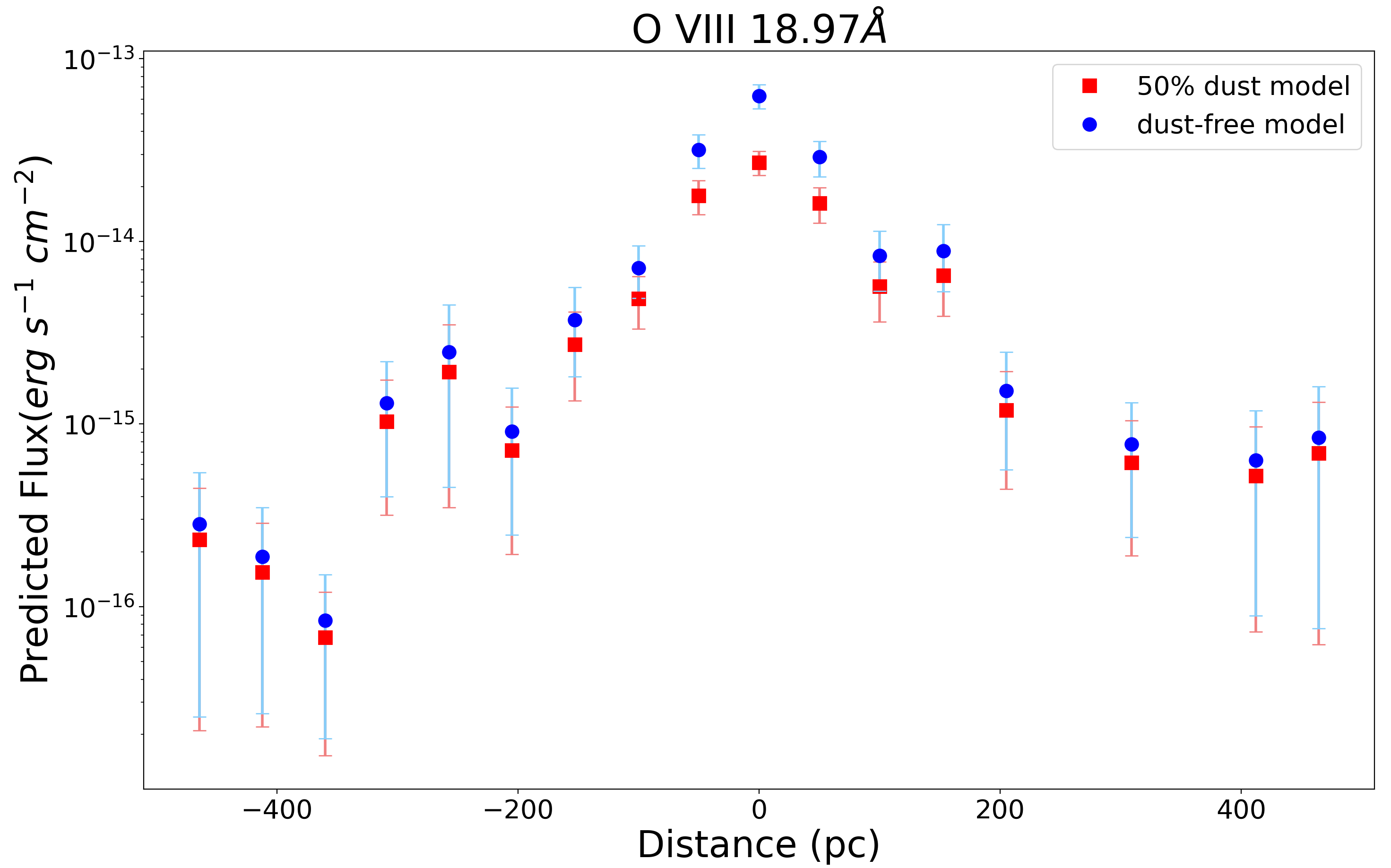}
 \end{minipage}\qquad 
 \begin{minipage}[b]{0.4\textwidth}
  \includegraphics[width=7cm, height=4.3cm]{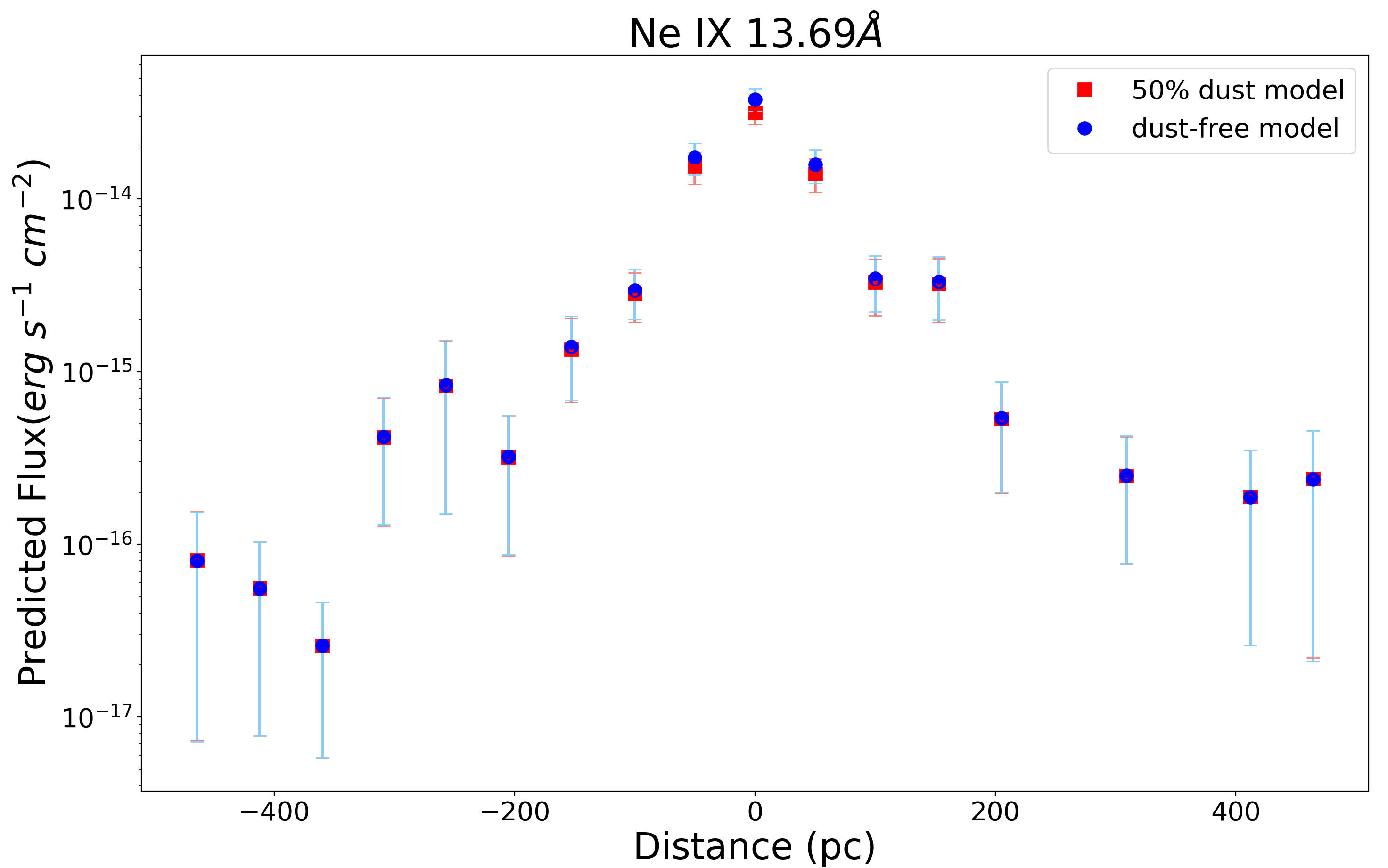}
\end{minipage}\qquad 
 \begin{minipage}[b]{0.4\textwidth}
  \includegraphics[width=7cm, height=4.3cm]{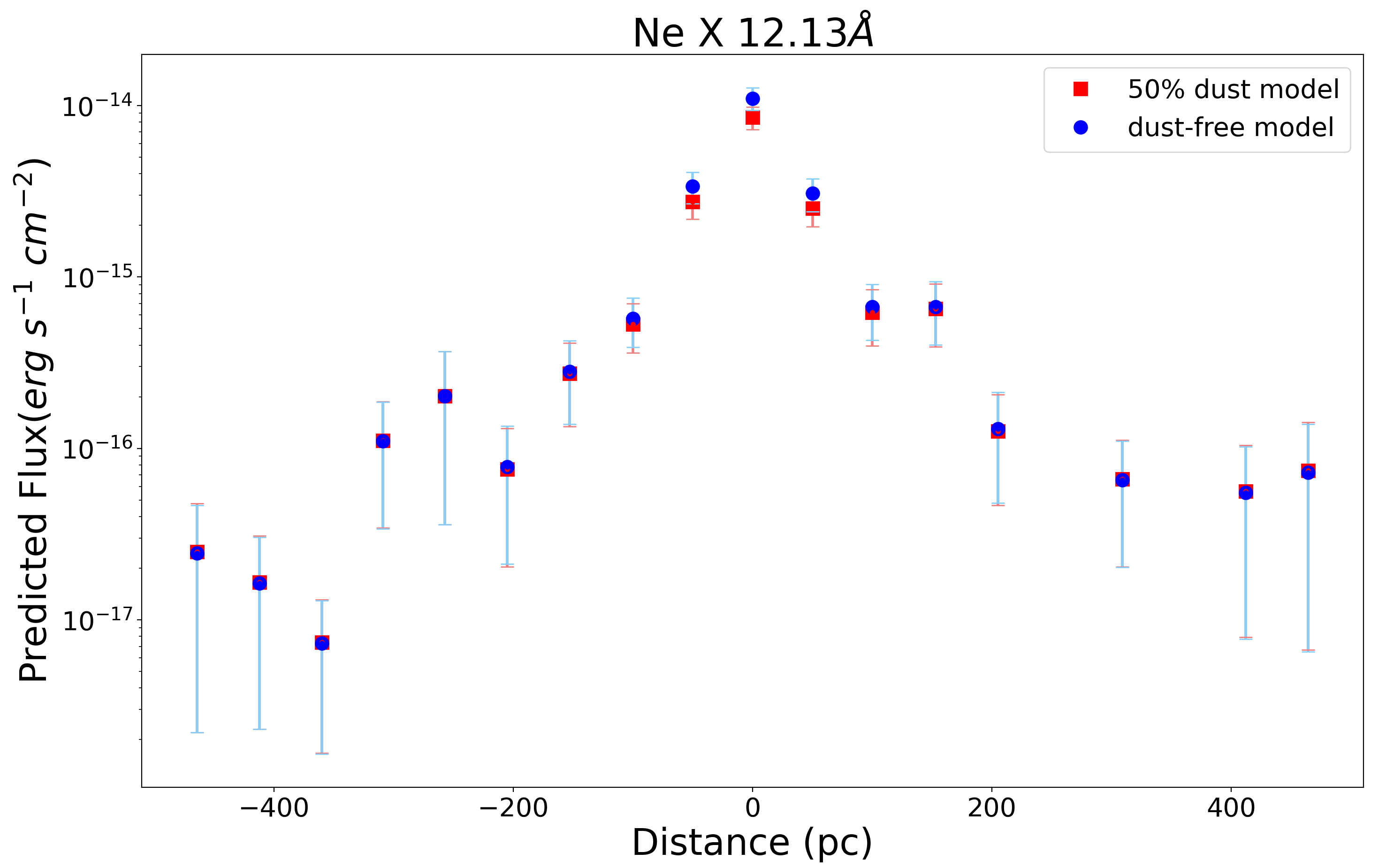}
 \end{minipage}\qquad

\caption{Computed radial flux distributions for X-ray lines O~VII 22.1\AA, O~VIII 18.97\AA, Ne~IX 13.69\AA, and Ne~X 12.13\AA. The fluxes are obtained using the analysis described in section \ref{sec:models}. The dust-free models are plotted in red and the 0.5 ISM models are plotted in blue. The models assume $\delta r/r$ $<$ 1 or $\delta r$ fixed at 50pc, where $\delta r$ is the deprojected width for each extraction bin. The uncertainties were obtained based on the S/N of the fluxes in the 0th order image from \citealt{kraemer2020a}.}
\label{fig:fluxes4}
\end{figure*}

\citet{kraemer2020a} scaled $Q$ using an average UV flux at 1350 \AA~ observed over the last 7 years, and computed the value for $Q$ using the assumed SED and a distance of 15 Mpc. They obtained a value of $Q = 3\times 10^{53}$ photons ${\rm s^{-1}}$ and a bolometric luminosity L$_{bol}$ = 1.4$\times10^{44}$ ${\rm erg~s^{−1}}$ or 0.03 of the Eddington Luminosity. \par 

In addition, the elemental abundances in the NLR of NGC 4151 were initially determined in \citealt{kraemer2000b}, and based on recent estimates of solar elemental abundances \citep[e.g.,][]{asplund2005a}, they correspond to roughly 1.4x solar. We assume the same values for the present paper, as follows (in logarithm, relative to H, by number):
He=~-$1.00$, C=~-$3.47$, N=~-$3.92$, O=~-$3.17$, Ne=~-$3.96$, Na=~-$5.76$, Mg=~-$4.48$, Al=~-$5.55$, Si=~-$4.51$, P=~-$6.59$, S=~-$4.82$, Ar=~-$5.60$, Ca=~-$5.66$, Fe=~-$4.4$, and Ni=~-$5.78$.\par 

We also generate photoionisation models that account for internal dust, assuming grain abundances of 50\% of those determined for the interstellar medium (ISM) \citep[e.g.,][]{mathis1977a}, with the depletions from gas phase scaled accordingly \citep[e.g.,][]{snow1996a}, i.e., C=~-$3.64$, O=~-$3.29$, Mg=~-$4.78$, Al=~-$6.45$, Si=~-$4.81$, Ca=~-$5.96$, Fe=~-$4.70$, and Ni=~-$6.08$. These depletions are consistent with more recent studies \citep[e.g.,][]{mehdipour2018a}. The grain abundances of 50\% is the same as that adopted by \citet{kraemer2000b} in their photoionisation study of the NLR of NGC 4151.\par

If the X-ray gas was created from the expansion of [O~III] clouds, as discussed in \citealt{trindadefalcao2021b}, we expect it to possess internal dust, since the predicted dust temperatures of the X-ray models are too low to be able to destroy the dust grains \citep[e.g.,][]{barvainis1987a}. On the other hand, the destruction of dust grains is possible as a result of shocks. If the X-ray gas is entraining clouds of [O~III]-emitting gas \citep[][]{trindadefalcao2021b} and effectively "sweeping" them up, the shocks that occur during this process could be enough to shatter the dust grains within the gas \citep[e.g.,][]{jones1997a}. Our assumption of lower than ISM grain abundance is consistent with this scenario.\par

  \begin{table}
\footnotesize
  \centering
\begin{tabular}[b]{|l|c|c|r|}
\hline
 \textbf{Distance$^{a}$} & \textbf{log$U$}  & \textbf{log $n_{H}^{a}$} & \textbf{log $N_{H}^{b}$}  \\
  & & ${\rm cm^{-3}}$ & ${\rm cm^{-2}}$ \\
\hline
12 pc & 0.05 & 2.71 & 22.28\\
50 pc & -0.17 & 1.69 & 21.88\\
100 pc & -0.27 & 1.19 & 21.38\\
153 pc & -0.34 & 0.89 & 21.06 \\
205 pc & -0.38 & 0.68 & 20.81 \\
257 pc & -0.42 & 0.52 & 20.71 \\
309 pc & -0.45 & 0.39 & 20.58 \\
360 pc & -0.47 & 0.28 & 20.27\\
412 pc & -0.49 & 0.18 & 20.37\\
464 pc & -0.50 & 0.09 & 20.28\\

\hline
\end{tabular}
\caption{\textbf{Cloudy Model Parameters.} The models follow the same parameters as the medium ionisation component from \citealt{kraemer2020a}\\
\small $^{a}$ Based on density law $n_{H}$ $\propto$ $r^{-1.65}$\\
$^{b}$ Assuming $\delta r/r$ $<$ 1 or $\delta r$ fixed at 50 pc, where $\delta r$ is the deprojected width for each extraction bin; for a detailed explanation, see \citealt{kraemer2020a}. }
\label{tab:models}
\end{table}

\subsection{Radial Flux Distributions}

In their study, \citet{kraemer2020a} calculated the emitting area for the X-ray emitting gas. They were able to do this calculation only for Ne~IX 13.69\AA~ and Ne~X 12.13\AA, since the loss of soft-Xray sensitivity for ACIS made it hard to accurately measure the O~VII 22.1\AA~and O~VIII 18.97\AA~fluxes. To construct the radial flux profiles associated with each emission-line, we redo their analysis, and include in our study other X-ray line profiles, e.g., O~VII 22.1\AA~and O~VIII 18.97\AA. In addition, in our study, we only generate photoionisation models for the medium ionisation component from \citealt{kraemer2020a}. The reason is that the highest ionisation models did not predict significant fluxes for the footprint lines. The complete set of parameters used to generate our models is listed in Table \ref{tab:models}.  \par

Using the Cloudy predictions for the X-ray footprint lines, and the emitting areas calculated in \citealt{kraemer2020a}, we also construct radial flux profiles for the footprint lines, such as [Fe~X] 6375\AA~and [Si~X] 1.43$\mu$m. Our results are shown in Figure \ref{fig:fluxes1} and \ref{fig:fluxes1_1}, for the optical/IR lines, and in Figure \ref{fig:fluxes4} for the X-ray lines. It is important to note that our radial flux profiles show two different sets of models: dust-free and 50\% dust. The major effect for the dusty models is the depletion of elements onto dust grains. For example, since sulfur and neon are not depleted from gas phase, the [S~X], Ne~IX and Ne~X do not show any significant differences, other than a slight increase at the points closest to the AGN. The resulting spatially-integrated fluxes for all the footprint and X-ray lines are listed in Table \ref{tab:fluxes}. \par 

It is important to note that a few of our predicted flux profiles, such as [Ca~VIII], and [Fe~VII] (see Figure \ref{fig:fluxes1}) show a higher predicted flux for the dusty models in the center bins. This is because the He~II zone is shallower when dust is present, which results in greater recombination into those ionisation states. However, the effect is not present for the more distant points, because N$_{H}$ is smaller (see Table \ref{tab:models}), hence the model integrations do not reach the end of the  He~II zone \citep[see discussion in][]{kraemer1986a}. \par

\section{Footprints of X-ray Outflows}
\label{sec:footprints}

\subsection{Observational Limits on the Footprint Lines}
\begin{figure}
  \centering
  \includegraphics[width=0.5\textwidth]{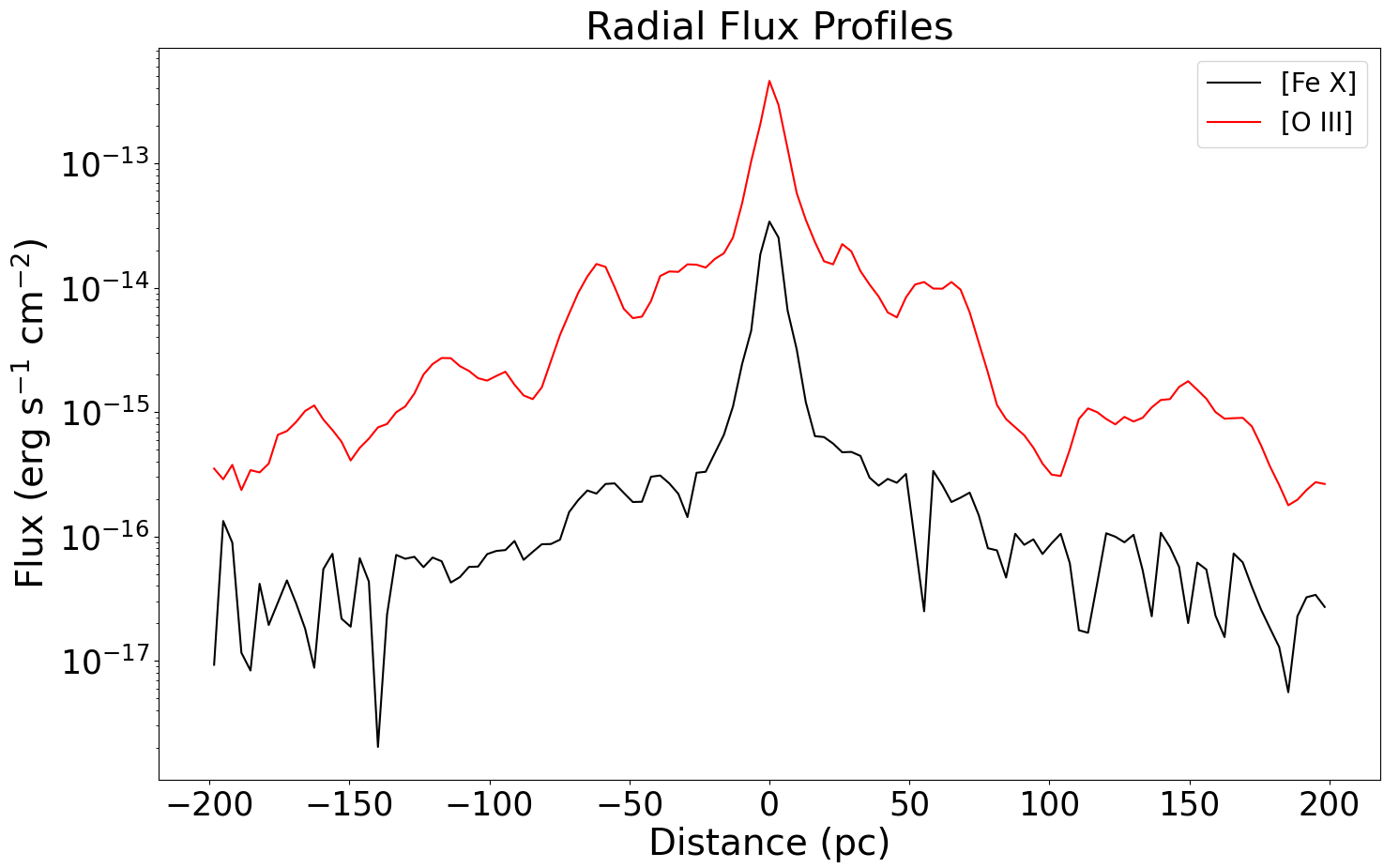}
 \caption{Spatial distribution for [Fe~X] (in black) and [O~III] (in red) in NGC 4151. The values were obtained from the analysis of the STIS G750L spectrum. The fluxes are per cross-dispersion pixel (0.1$\arcsec$ $\times$ 0.05$\arcsec$).}
\label{fig:spatial_dist}
\end{figure}

By analysing the archival STIS long-slit spectra, we are able to obtain the spatial [Fe~X] and [O~III] distribution in NGC 4151, as shown in Figure \ref{fig:spatial_dist}. Specifically, since the [Fe~X] $\lambda$6375 emission-line is contaminated by [O~I] $\lambda$6364, we obtain an intrinsic [Fe~X] brightness profile by subtracting 1/3 of the [O~I] $\lambda$6300 flux at each point along the slit. In addition, in order to account for the mass of [Fe~X] emitting gas outside of the area sampled by the STIS slit, we use an archival HST/WFPC2 [O~III] image of the NLR for the target \citep[][]{kraemer2020a}. We measure the [O~III] fluxes in regions of 0$\arcsec$.5x3$\arcsec$.0, centered at the nucleus, along a position angle of 140\degree. \par 

\begin{figure}
  \centering
  \includegraphics[width=0.5\textwidth]{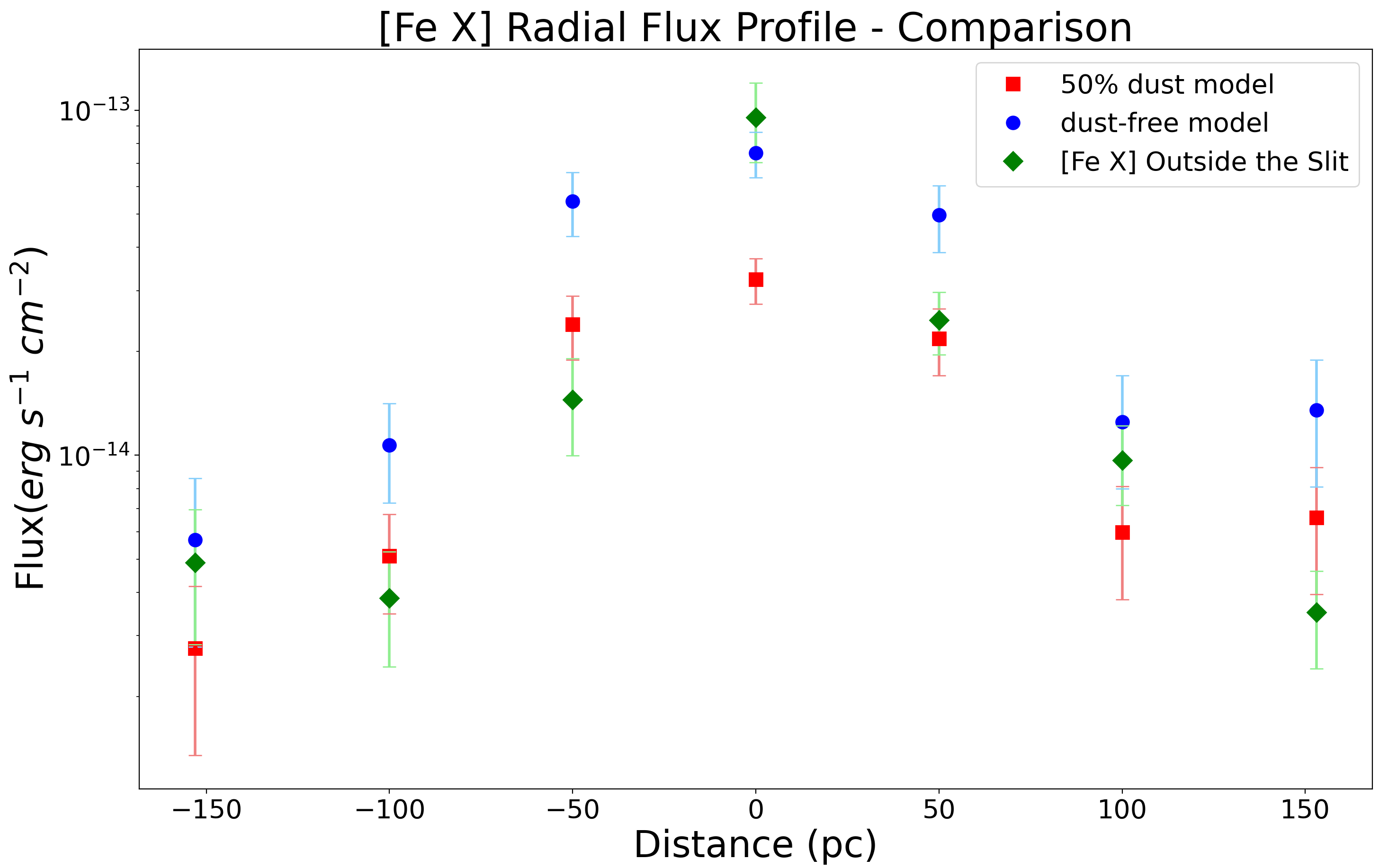}
\caption{A comparison between the results of our predicted radial flux profiles for the [Fe~X]-emitting gas (the blue points represent the dust-free model values and the red points represent the 50\% dust model values) and the results of the full [Fe~X] flux distribution (inside and outside the STIS slit). The uncertanties were obtained based on the S/N of the fluxes in the 0th order image from \citet{kraemer2020a}. Additionally, the flux profiles show lower fluxes for the negative distances. This is because the observed fluxes are asymmetric around the nucleus, which is not easily seen in Figure \ref{fig:spatial_dist}, since the slit did not cover the entire NLR.}
\label{fig:comparison}
\end{figure}

We then measure the ratio [Fe~X]/[O~III] inside the STIS slit, which allow us to obtain the full [Fe~X] flux distribution. In Figure \ref{fig:comparison}, we compare the results of our predicted [Fe~X] flux profiles (both for dust-free models and 50\% dust models) and the results of the full [Fe~X] flux distribution (inside and outside the slit). The total spatially-integrated [Fe~X] flux, i.e., the sum of the fluxes inside and outside the STIS slit, is $1.9^{+0.6}_{-0.2}\times10^{-13}~{\rm erg~s^{-1}~cm^{-2}}$. It is important to note here that the flux profiles show lower fluxes for the negative distances. This is because the observed fluxes are asymmetric around the nucleus, which is not easily seen in Figure \ref{fig:spatial_dist}, since the slit did not cover the entire NLR.\par 

In addition, based on the analysis of the STIS long-slit spectra, we are able to calculate an upper limit, integrated over 11 pixels (0.1$\arcsec$ $\times$ 0.55$\arcsec$), for the flux of the [Fe~XIV] 5303\AA~line $<$ $7.5\times 10^{-14}~{\rm erg~s^{-1}~cm^{-2}}$, which agrees with our predicted flux of $1.3\times 10^{-14}~{\rm erg~s^{-1}~cm^{-2}}$ (see Table \ref{tab:fluxes}). Considering that the brightness profile for [Fe~XIV] follows the same brightness profile as [Fe~X], we calculate an upper limit in the central bin 0.1$\arcsec$ $\times$ 0.05$\arcsec$ $<$ $2.6\times10^{-14}~{\rm erg~s^{-1}~cm^{-2}}$, which also agrees with our predicted value of $1.0\times10^{-14}~{\rm erg~s^{-1}~cm^{-2}}$. It is important to note here that the other optical footprint lines are below the detection limit of STIS. \par

\subsection{Predicting the Mass of the Extended Gas}
\label{sec:mass_prediction}

In order to recompute the mass of the extended X-ray gas in NGC 4151 based on our X-ray footprint analysis, we use our predicted [Fe~X] radial flux profile. It is important to note here that the methodology used in this study to estimate the mass of the extended X-ray gas in the target is different from the methodology used by \citet{kraemer2020a}. \citet{kraemer2020a} mapped the Ne~IX triplet (the resonance line, 13.50\AA, the intercombination line, 13.60\AA, and the forbidden line, 13.73\AA) and Ne~X~$\alpha$, instead of [Fe~X], and applied the model parameters to fit the 0th order emission-line profiles of these lines.\par 

\begin{figure}
  \centering
  \includegraphics[width=0.5\textwidth]{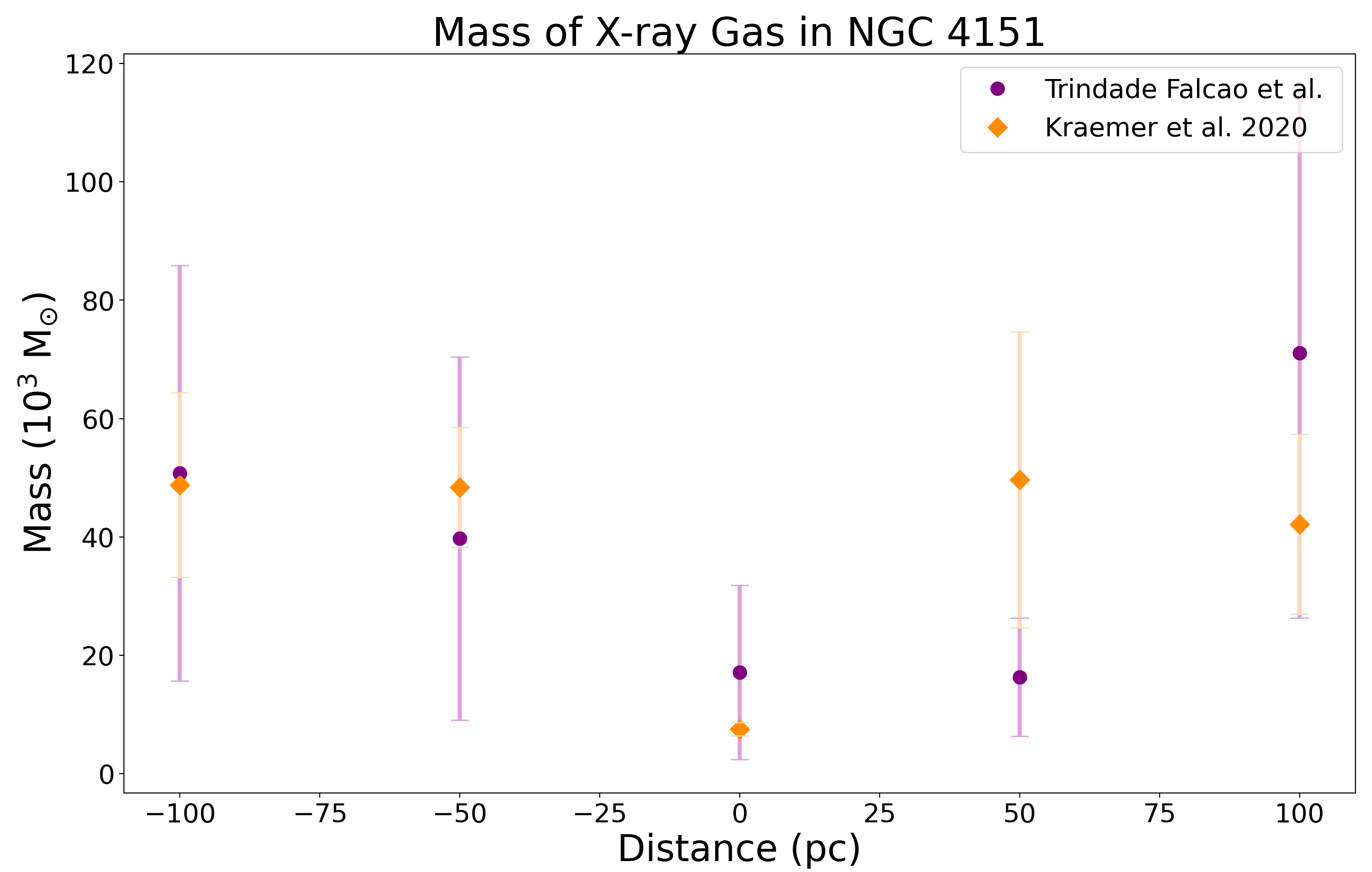}
\caption{A comparison between the results of our predicted masses for the X-ray-emitting gas (purple points) and the results of \citealt{kraemer2020a} (orange points). The uncertainties were obtained based on the S/N of the fluxes in the 0th order image from \citet{kraemer2020a}.}
\label{fig:comparison_masses}
\end{figure}

We calculate the masses in each bin by using the line luminosities and the Cloudy-predicted fluxes for the [Fe~X], as follows:

\begin{equation}
    M_{bin} = N_{H}\mu m_{p} \left(\frac{L_{[Fe X]}}{F_{{[Fe X]}_{c}}}\right)
\end{equation}

\noindent where $N_{H}$ is the hydrogen column density of each photoionisation model generated by \citet{kraemer2020a}, assumed to be the same as the column density modelled by Cloudy, $\mu$ is the mean mass per proton, which we assume to be 1.4 (consistent with roughly solar abundances), and $m_{p}$ is the mass of a proton. In order to get the mass of [Fe~X] the column density needs to be multiplied by an effective area. The term in parentheses gives the effective surface area of the emitting gas as seen by the observer. $F_{{[Fe X]}_{c}}$ is the [Fe~X] luminosity per ${\rm cm^{2}}$,
predicted by Cloudy, and $L_{[Fe X]}$ is the observed luminosity determined using the total flux. Specifically,

\begin{equation}
   L_{[Fe X]} = 4\pi D^{2} F_{[Fe X]_{total}}
\end{equation}

\noindent where $D$ is the distance to NGC 4151, and $F_{[Fe X]_{total}}$ is the total [Fe~X] flux.
Physically, the equation for $M_{slit}$ determines the area of the emitting clouds through the ratio of the luminosity and flux, and then multiplies this by the column density to yield the total number of particles, which, when multiplied by the mean mass per particle, gives the total ionised mass. The total mass of X-ray-emitting gas is $7.8(\pm 2.1)\times 10^{5}~M_{\odot}$, with the uncertainties being added in quadrature for each bin, and the mass profile for the target is shown in Figure \ref{fig:comparison_masses} (purple points). We compare our results to the results of \citealt{kraemer2020a}, as shown in Figure \ref{fig:comparison_masses} (orange points), which found that the total mass of X-ray gas within the outflow regions for NGC 4151 is $5.4(\pm 1.1)\times 10^{5}~M_{\odot}$. \par 

Even though our results for the total mass of X-ray-emitting gas are very promising, due to the poor resolution of the STIS grating used for this observation, along with a low S/N, we are not able to obtain any accurate kinematic information for [Fe~X]. This means that we cannot compute the outflow rates for NGC 4151, and a comparison to the values for $\dot{M}$ found by \citet{kraemer2020a}, i.e., $\dot{M}$ = 1.8M$_{\odot}$ ${\rm yr^{-1}}$, is not possible given the available STIS low-dispersion spectrum.

  \begin{table*}
\normalsize
  \centering
\begin{tabular}[b]{|l|c|c|c|c|c|c|}
\hline
 \textbf{Ion} & \textbf{Wavelength}  & \textbf{IP} & \textbf{Peak log$U^{*}$} & \textbf{Predicted Flux df$^{a}$} & \textbf{Predicted Flux d$^{b}$} & \textbf{Observed Flux} \\
  & & (eV) & & $({\rm 10^{-16}~erg~s^{-1}~cm^{-2}})$ & $({\rm 10^{-16}~erg~s^{-1}~cm^{-2}})$ & $({\rm 10^{-16}~erg~s^{-1}~cm^{-2}})$\\
\hline
\boldsymbol{${\rm[Al~IX]}$} & 2.04$\mu$m & 284 & 0.0 & 51.8 ($\pm$4.8) & 6.9 ($\pm$0.6) & 23 ($\pm$16)$^{c}$ \\
${\rm[Ca~VIII]}$ & 2.32$\mu$m & 127 & -0.75 & 3.9 ($\pm$0.6) & 5.3 ($\pm$0.5) & 151 ($\pm$23)$^{c}$ \\
${\rm[Fe~VII]}$ & 6087\AA & 99 & -1.0 & 15.4 ($\pm$2.0) & 26.5 ($\pm$2.5) & -\\
\boldsymbol{${\rm[Fe~X]}$} & 6375\AA & 234 & 0.0 & 2360 ($\pm$215.0) & 1050 ($\pm$96.2) & 1900$^{d}$\\
\boldsymbol{${\rm[Fe~XI]}$} & 7892\AA & 262 & 0.25 & 921 ($\pm$87.7) & 302 ($\pm$27.7) & -\\
\boldsymbol{${\rm[Fe~XIV]}$} & 5303\AA & 361 & 0.5 & 130 ($\pm$15.6) & 34.0 ($\pm$3.9) & $<$750$^{d}$\\
\boldsymbol{${\rm[Mg~VII]}$} & 5.50$\mu$m & 186 & -0.75 & 35.2 ($\pm$4.2) & 37.1 ($\pm$3.4) & -\\
\boldsymbol{${\rm[Mg~VIII]}$} & 3.02$\mu$m & 225 & -0.25 & 251 ($\pm$24.8) & 174 ($\pm$15.9) & -\\
\boldsymbol{${\rm[S~IX]}$} & 1.25$\mu$m & 328 & -0.25 & 22.0 ($\pm$2.2) & 31.1 ($\pm$2.9) & 397 ($\pm$23)$^{c}$\\
\boldsymbol{${\rm[Si~VII]}$} & 2.48$\mu$m & 205  & -1.0 & 14.6 ($\pm$ 1.7) & 16.9 ($\pm$1.5) & -\\
\boldsymbol{${\rm[Si~IX]}$} & 3.55$\mu$m  & 303 & 0.0 & 319 ($\pm$33.0) & 195 ($\pm$17.3) & -\\
\boldsymbol{${\rm[Si~X]}$} & 1.43$\mu$m & 351 & 0.25 & 559 ($\pm$51.9) & 214 ($\pm$19.5) & 377 ($\pm$30)$^{c}$\\
\textbf{O~VII f} & 22.1\AA & 138 & -0.5 & 3940 ($\pm$359.0) & 3230 ($\pm$293.0) & 3400($\pm$769)$^{e}$\\
\textbf{O~VIII} \boldsymbol{$\alpha$} & 18.97\AA & 739 & 0.25 & 610 ($\pm$146.0) & 880 ($\pm$79.5) & 1089($\pm$192)$^{e}$\\
\textbf{Ne~IX i} & 13.69\AA & 239 & 0.0 & 851 ($\pm$79.4) & 747 ($\pm$69.1) & 1041($\pm$96)$^{e}$\\
\textbf{Ne~X} \boldsymbol{$\alpha$} & 12.13\AA & 1195 & 0.75 & $^{**}$204 ($\pm$19.8) & 166 ($\pm$15.8) & 480($\pm$48)$^{e}$\\

\hline
\end{tabular}
\caption{\textbf{Predicted Spatially-Integrated Fluxes.} In boldface are the emission-lines that we define as footprint lines, i.e., lines from ions with IP $\geq$ 138 eV.\\
\small $^{a}$ Dust-free models\\
$^{b}$ 50\% dust models\\
$^{c}$ Observed fluxes from \citealt{rodriguez-ardila2011a}\\
$^{d}$ Observed fluxes from this study\\
$^{e}$ Observed fluxes from \citealt{kraemer2020a}\\
$^{*}$ The values of log$U$ are to the nearest 0.25dex. All the models are for log N$_{H}$=21.5.\\
$^{**}$Most of the X-ray emission accounted for this line comes from the high ionisation components modeled by \citet{kraemer2020a}. However, since these models do not significantly contribute to the footprint lines emission we decide not to include them in this study.}
\label{tab:fluxes}
\end{table*}

\section{Discussion}
\label{sec:discussion_1}

\begin{figure}
  \centering
  \includegraphics[width=0.5\textwidth]{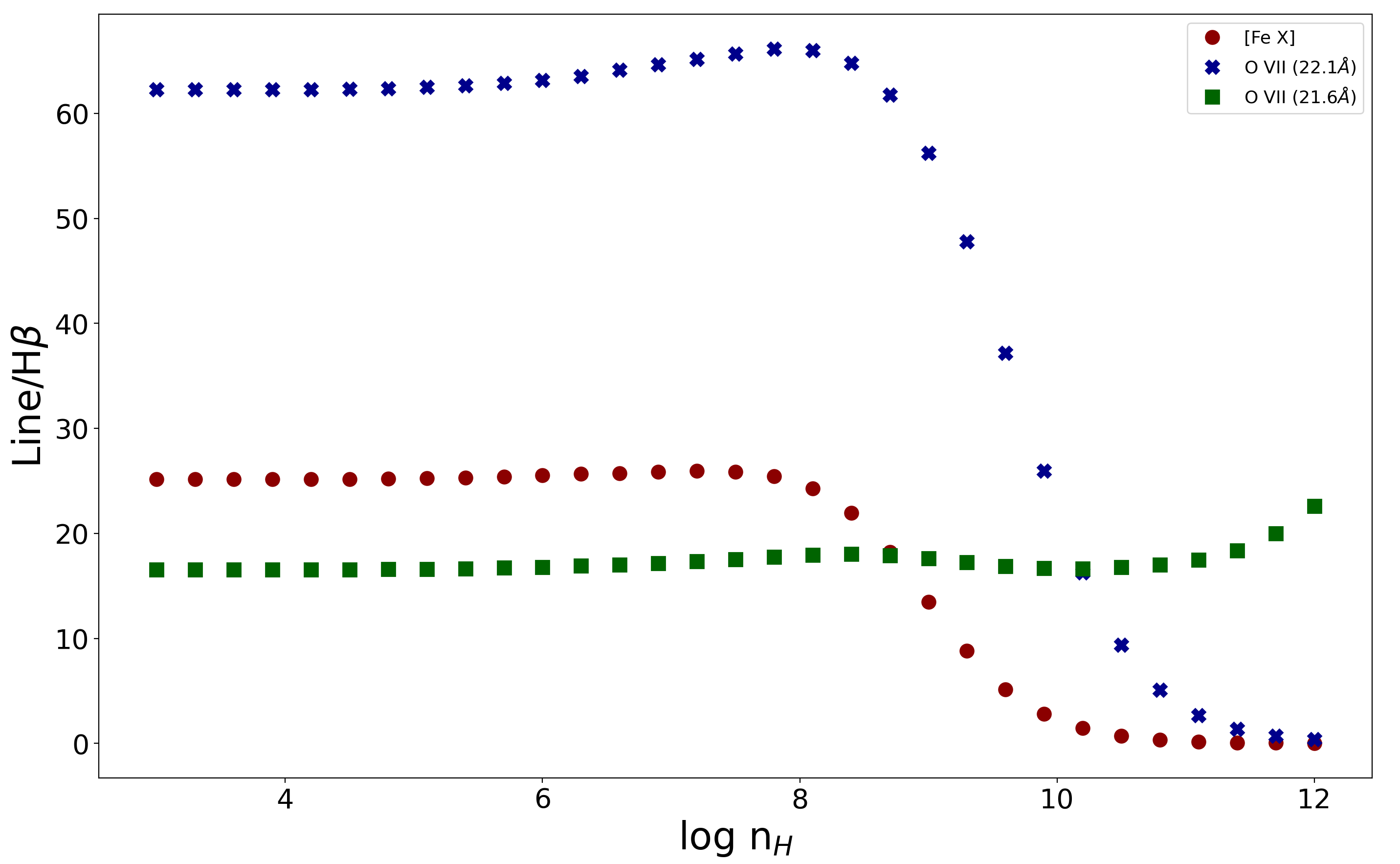}
\caption{Predicted [Fe~X]/H${\beta}$ (in red), O~VII -f/H${\beta}$ (in blue), and O~VII -r/H${\beta}$ (in green) ratios for NGC 4151. The model assumptions are: abundances (1.4x solar), SED from \citealt{kraemer2020a}, constant column density N$_{H}$ = 10$^{21.5}$ ${\rm cm^{-2}}$, and ionisation parameter log$U$ = 0.0.}
\label{fig:ratios}
\end{figure}

In Figure \ref{fig:ratios} we present the Cloudy-predicted [Fe~X]/H${\beta}$ (in red), O~VII~(22.1\AA)/H${\beta}$ (in blue), and O~VII~(21.6\AA)/H${\beta}$ (in green) ratios for NGC 4151, for the 50\% dust fraction models. It is possible to see that when densities are very high, i.e.,  $\gtrsim$ 10$^{8.0}$ ${\rm cm^{-3}}$, the [Fe~X]/H${\beta}$ ratio starts to drop, which means that the [Fe~X] 6375\AA~line starts to become collisionally suppressed. However the O~VII~(22.1\AA)/H${\beta}$ ratio also drops with increasing density, hence [Fe~X] still tracks the O~VII 22.1\AA~emission. On the other hand, the O~VII 21.6\AA~line is not suppressed over this range in density. Therefore, there could be a contribution to the X-ray emitting gas from denser regions, close to the AGN. However, for NGC 4151, $Q \approx$ $3\times 10^{53}~{\rm photons~s^{-1}}$ \citep[e.g.,][]{kraemer2020a}, so gas characterised by log$U$ = 0.0 with densities of $\approx$ 10$^{9}~{\rm cm^{-3}}$ would lie at $r \approx$ $2.8\times10^{16}$ cm from the central SMBH, which is just at the outer edge of the BLR (about 10 light-days) \citep[e.g.,][]{bentz2013a}. The fact that our predicted values for the mass of X-ray-emitting gas in NGC 4151 agree very well with the measured values by \citet{kraemer2020a} (see Figure \ref{fig:comparison_masses}) tells us that the contribution from X-ray gas that is located in the BLR of this AGN is negligible. This result is consistent with the findings of \citet{crenshaw2007a}, which shows that the mass outflows from the dense inner region of the AGN do not contribute significantly to the total gas in the NLR of NGC 4151, and the gas located in this region, i.e., in the NLR, is likely accelerated in-situ.  \par

As discussed in Section \ref{sec:models}, internal dust can have a significant effect on how we interpret our results. As shown in Figure \ref{fig:fluxes1} and \ref{fig:fluxes1_1}, the presence of dust within the outflowing gas can have a more pronounced effect in some lines, e.g., [Fe~X] 6375\AA, compared to others, e.g., [S~IX] 1.25$\mu$m. The fact that the [S~IX] fluxes do not seem to be heavily affected by the presence of dust within the gas, as shown in Figure \ref{fig:fluxes1_1}, tells us that this line might be a good choice of footprint line when the percentage of dust within the gas is unknown (but see below). On the other hand, for lines like [Fe~X] 6375\AA, and [Mg~VIII] 3.02$\mu$m it is important to have a good estimate of the amount of dust present in the gas, since their fluxes are strongly affected by depletion of these element onto dust grains. For instance, if one uses only the [Fe~X] 6375\AA~line to predict the emission of X-ray gas, but underestimates the amount of dust within the gas, the estimated X-ray mass could be overestimated, since the dusty models predict lower fluxes, as shown in Figure \ref{fig:fluxes1}. \par 

In addition, \citet{rodriguez-ardila2011a} measured the fluxes of IR footprint lines for NGC 4151 in their study of near-infrared coronal line spectrum of nearby AGN. As shown in Table \ref{tab:fluxes}, our predicted fluxes for [Al~IX] 2.04$\mu$m and [Si~X] 1.43$\mu$m agree well with the results of \citet{rodriguez-ardila2011a}. The under-prediction of [Ca~VIII] 2.32$\mu$m is consistent with the fact that Ca~VIII peaks at much lower ionisation than our X-ray models (see Figure \ref{fig:cloudy}). Regarding the discrepancy between our measured flux for the [S~IX] 1.25$\mu$m and the observed flux from \citealt{rodriguez-ardila2011a}, we believe that our underpredicted value could be due to collision strength of this transition \citep[e.g.,][]{zhang2002a}, or due to an unmodeled process, such as photo-excitation/fluorescence.\par 

It is important to note that, in this study, we use [Fe~X] as a tracer for X-ray outflows based on the characteristics of NGC 4151, i.e., the fact that this target is not a very powerful AGN. For AGN with more highly ionised gas in their NLR, other lines would be a better choice, such as [Si~X] and [Fe~XIV], as shown in Figure \ref{fig:cloudy}. For instance, based on our predicted flux and observed flux for [Fe~XIV] 5303\AA (see Table \ref{tab:fluxes}), there is not a large amount of [Fe~XIV] in NGC 4151 and this is because there is only a small amount of high ionisation gas in the NLR of this target. For more powerful targets, e.g., quasars, the use of [Fe~XIV] as a tracer might be preferable, since these objects are more luminous and powerful than NGC 4151. \par

\section{Conclusions}
\label{sec:discussion}

Based on our analysis derived from Cloudy photoionisation models, we present a new method that can be used to model X-ray outflows. Our conclusions are as follows:\par 

1. There are ions that peak over the same range in ionisation parameter as the H and He-like ions and can produce detectable optical/IR emission lines, which we call "footprint lines". These lines can be detected and spatially-resolved, e.g., via \textit{HST} or \textit{JWST} observations and, therefore, can be used to accurately measure the kinematics and the dynamics of the extended X-ray gas. For instance, footprint lines such as [Al~IX], [Si~VII], [Si~IX], and [Si~X] can be observed with \textit{JWST}, and, for NIRSpec IFU, where most of the footprint lines are, the spectral resolving power R$\sim$100, 500, 1000s (depending  on  the  mode) corresponds to a velocity resolution of <$\sim$100-300 km/s, which will allow us to obtain good kinematic information on the targets and, subsequently, be able to fully characterise the X-ray outflows.\par 

2. We show that the models based on \textit{Chandra} data from \citealt{kraemer2020a} predict strong footprint lines, which is consistent with the \textit{HST} data. We also develop flux profiles to characterise such lines. \par 

3. By analysing the STIS long-slit spectrum and the [O~III] image for NGC 4151, we are able to derive the measured flux profile for [Fe~X] 6375\AA~within similar extraction bins. \par 

4. We compare our model predictions to the observed flux profile for [Fe~X] obtained from the analysis of the STIS G750L spectrum. Our values are in very good agreement with the measured values, which confirm the reliability of our approach. We also compare our model predictions for the IR lines to the observed fluxes from \citealt{rodriguez-ardila2011a}, which show very good agreement for the [Al~IX] 2.04$\mu$m and [Si~X] 1.43$\mu$m (see Section \ref{sec:discussion_1} for caveat on predicted IR footprint lines).\par 

5. We confirm our methodology by determining the mass of X-ray gas using our models predictions and the observed flux profile for the [Fe~X]. As shown in Figure \ref{fig:comparison_masses}, our values are in very good agreement with the values found by \citet{kraemer2020a}, which confirms that our method can accurately produce mass profiles for the X-ray gas. \par 

\medskip
 
Even though we are able to produce accurate mass profiles for the X-ray gas using our method of "footprint lines", we are not able to get accurate kinematic information for these lines from the STIS data. The poor resolution of the STIS G750L grating, combined with low S/N, does not allow us to obtain the necessary kinematic information to compute the values for outflow rates of this target. It would be interesting to obtain observations with higher spectral resolution for NGC 4151, which would allow us to compute the predicted mass outflow rates and again test our proposed methodology. Another possibility is to apply this method on other AGN, such as NGC 1068, for which the combination of X-ray spectra and medium resolution STIS long-slit spectra exist. In addition, it is important to note that, even though we do not need the kinematic profiles derived from X-ray observations to apply our method, we would still need X-ray spectra to compare with photoionisation models in order to fully characterise the ionisation structure of these outflows.

\section*{Acknowledgements}

The material is based upon work supported by NASA under award number 80GSFC21M0002. Basic research at the Naval Research Laboratory is funded by 6.1 base funding.\par 
This research has made use of the NASA/IPAC Extragalactic Database (NED), which is operated by the Jet Propulsion Laboratory, California Institute of Technology, under contract with the National Aeronautics and Space Administration. This paper used the photoionisation code Cloudy, which can be obtained from http://www.nublado.org. We thank Gary Ferland and associates, for the maintenance and development of Cloudy. 

\section*{Data Availability}

Based on observations made with the NASA/ESA Hubble Space Telescope, and available from the Hubble Legacy Archive, which is a collaboration between the Space Telescope Science Institute (STScI/NASA), the Space Telescope European Coordinating Facility (ST-ECF/ESAC/ESA) and the Canadian Astronomy Data Centre (CADC/NRC/CSA).


\bibliographystyle{mnras}
\bibliography{anna_bibliography} 

\begin{thebibliography}{}
\makeatletter
\relax
\def\mn@urlcharsother{\let\do\@makeother \do\$\do\&\do\#\do\^\do\_\do\%\do\~}
\def\mn@doi{\begingroup\mn@urlcharsother \@ifnextchar [ {\mn@doi@}
  {\mn@doi@[]}}
\def\mn@doi@[#1]#2{\def\@tempa{#1}\ifx\@tempa\@empty \href
  {http://dx.doi.org/#2} {doi:#2}\else \href {http://dx.doi.org/#2} {#1}\fi
  \endgroup}
\def\mn@eprint#1#2{\mn@eprint@#1:#2::\@nil}
\def\mn@eprint@arXiv#1{\href {http://arxiv.org/abs/#1} {{\tt arXiv:#1}}}
\def\mn@eprint@dblp#1{\href {http://dblp.uni-trier.de/rec/bibtex/#1.xml}
  {dblp:#1}}
\def\mn@eprint@#1:#2:#3:#4\@nil{\def\@tempa {#1}\def\@tempb {#2}\def\@tempc
  {#3}\ifx \@tempc \@empty \let \@tempc \@tempb \let \@tempb \@tempa \fi \ifx
  \@tempb \@empty \def\@tempb {arXiv}\fi \@ifundefined
  {mn@eprint@\@tempb}{\@tempb:\@tempc}{\expandafter \expandafter \csname
  mn@eprint@\@tempb\endcsname \expandafter{\@tempc}}}

\bibitem[\protect\citeauthoryear{{Asplund}, {Grevesse}  \& {Sauval}}{{Asplund}
  et~al.}{2005}]{asplund2005a}
{Asplund} M.,  {Grevesse} N.,   {Sauval} A.~J.,  2005, in COSMIC ABUNDANCES as
  Records of Stellar Evolution and Nucleosynthesis in honor of David L.
  Lambert. Astronomic Society of the Pacific, p.~25

\bibitem[\protect\citeauthoryear{{Barvainis}}{{Barvainis}}{1987}]{barvainis1987a}
{Barvainis} R.,  1987, The Astrophysical Journal, 320, 537

\bibitem[\protect\citeauthoryear{{Begelman}}{{Begelman}}{2004}]{begelman2004a}
{Begelman} M.~C.,  2004, Carnegie Observatories Astrophysics Series, 1, 374

\bibitem[\protect\citeauthoryear{Bentz et~al.,}{Bentz
  et~al.}{2013}]{bentz2013a}
Bentz M.~C.,  et~al., 2013, \mn@doi [The Astrophysical Journal]
  {10.1088/0004-637x/767/2/149}, 767, 149

\bibitem[\protect\citeauthoryear{{Bianchi}, {Chiaberge}, {Evans}, {Guainazzi},
  {Baldi}, {Matt}  \& {Piconcelli}}{{Bianchi} et~al.}{2010}]{bianchi2010a}
{Bianchi} S.,  {Chiaberge} M.,  {Evans} D.,  {Guainazzi} M.,  {Baldi} R.,
  {Matt} G.,   {Piconcelli} E.,  2010, Monthly Notices of the Royal
  Astronomical Society, 405, 553

\bibitem[\protect\citeauthoryear{{Crenshaw} \& {Kraemer}}{{Crenshaw} \&
  {Kraemer}}{2007}]{crenshaw2007a}
{Crenshaw} D.~M.,  {Kraemer} S.~B.,  2007, The Astrophysical Journal, 659, 250

\bibitem[\protect\citeauthoryear{{Crenshaw}, {Fischer}, {Kraemer}  \&
  {Schmitt}}{{Crenshaw} et~al.}{2015}]{crenshaw2015a}
{Crenshaw} D.~M.,  {Fischer} T.~C.,  {Kraemer} S.~B.,   {Schmitt} H.~R.,  2015,
  The Astrophysical Journal, 799, 83

\bibitem[\protect\citeauthoryear{{Di Matteo}, {Springel}  \& {Hernquist}}{{Di
  Matteo} et~al.}{2005}]{dimatteo2005a}
{Di Matteo} T.,  {Springel} V.,   {Hernquist} L.,  2005, Nature, 433, 604

\bibitem[\protect\citeauthoryear{{Ferland} et~al.,}{{Ferland}
  et~al.}{2017}]{ferland2017a}
{Ferland} G.~J.,  et~al., 2017, Revista Mexicana de Astronomia y Astrofisica,
  49, 1379

\bibitem[\protect\citeauthoryear{{Fischer} et~al.,}{{Fischer}
  et~al.}{2017}]{fischer2017a}
{Fischer} T.~C.,  et~al., 2017, The Astrophysical Journal, 834, 30

\bibitem[\protect\citeauthoryear{{Fischer} et~al.,}{{Fischer}
  et~al.}{2018}]{fischer2018a}
{Fischer} T.~C.,  et~al., 2018, The Astrophysical Journal, 856, 102

\bibitem[\protect\citeauthoryear{{Gebhardt} et~al.,}{{Gebhardt}
  et~al.}{2000}]{gebhardt2000a}
{Gebhardt} K.,  et~al., 2000, The Astrophysical Journal, 539, L13

\bibitem[\protect\citeauthoryear{{Gonzalez-Martin}, {Acosta-Pulido}, {Perez
  Garcia}  \& {Ramos Almeida}}{{Gonzalez-Martin} et~al.}{2010}]{gonzales2010a}
{Gonzalez-Martin} O.,  {Acosta-Pulido} J.~A.,  {Perez Garcia} A.~M.,   {Ramos
  Almeida} C.,  2010, The Astrophysical Journal, 723, 1748

\bibitem[\protect\citeauthoryear{{Hopkins} \& {Elvis}}{{Hopkins} \&
  {Elvis}}{2010}]{hopkins2010a}
{Hopkins} P.~F.,  {Elvis} M.,  2010, Monthly Notices of the Royal Astronomical
  Society, 401, 7

\bibitem[\protect\citeauthoryear{{Jones}, {Tielens}  \& {Hollenbach}}{{Jones}
  et~al.}{1997}]{jones1997a}
{Jones} A.~P.,  {Tielens} A. G. G.~M.,   {Hollenbach} D.~J.,  1997, ApJ, 469,
  740

\bibitem[\protect\citeauthoryear{{Kallman}, {Evans}, {Marshall}, {Canizares},
  {Longinotti}  \& {Schulz}}{{Kallman} et~al.}{2014}]{kallman2014a}
{Kallman} T.,  {Evans} D.~A.,  {Marshall} H.,  {Canizares} C.,  {Longinotti}
  A.~{Nowak} M.,   {Schulz} N.,  2014, The Astrophysical Journal, 780, 121

\bibitem[\protect\citeauthoryear{{Kraemer} \& {Harrington}}{{Kraemer} \&
  {Harrington}}{1986}]{kraemer1986a}
{Kraemer} S.~B.,  {Harrington} J.~P.,  1986, The Astrophysical Journal, 307,
  478

\bibitem[\protect\citeauthoryear{{Kraemer}, {Crenshaw}, {Hutchings}, {Gull},
  {Kaiser}, {Nelson}  \& {Weistrop}}{{Kraemer} et~al.}{2000}]{kraemer2000b}
{Kraemer} S.~B.,  {Crenshaw} D.~M.,  {Hutchings} J.~B.,  {Gull} T.~R.,
  {Kaiser} M.~E.,  {Nelson} C.~H.,   {Weistrop} D.,  2000, The Astrophysical
  Journal, 531, 278

\bibitem[\protect\citeauthoryear{{Kraemer}, {Ferland}  \& {Gabel}}{{Kraemer}
  et~al.}{2004}]{kraemer2004a}
{Kraemer} S.,  {Ferland} G.~J.,   {Gabel} J.~R.,  2004, \apj, 604, 556

\bibitem[\protect\citeauthoryear{{Kraemer}, {Turner}, {Couto}, {Crenshaw},
  {Schmitt}, {Revalski}  \& {Fischer}}{{Kraemer} et~al.}{2020}]{kraemer2020a}
{Kraemer} S.~B.,  {Turner} T.~J.,  {Couto} J.~D.,  {Crenshaw} D.~M.,  {Schmitt}
  H.~R.,  {Revalski} M.,   {Fischer} T.~C.,  2020, Monthly Notices of the Royal
  Astronomical Society, 493, 3893

\bibitem[\protect\citeauthoryear{{Lamperti} et~al.,}{{Lamperti}
  et~al.}{2017}]{lamperti2017a}
{Lamperti} I.,  et~al., 2017, Monthly Notices of the Royal Astronomical
  Society, 467, 540

\bibitem[\protect\citeauthoryear{{Landt}, {Ward}, {Steenbrugge}  \&
  {Ferland}}{{Landt} et~al.}{2015}]{landt2015a}
{Landt} H.,  {Ward} M.~J.,  {Steenbrugge} K.~C.,   {Ferland} G.~J.,  2015,
  Monthly Notices of the Royal Astronomical Society, 449, 3795

\bibitem[\protect\citeauthoryear{{Maksym} et~al.,}{{Maksym}
  et~al.}{2019}]{maksym2019a}
{Maksym} W.~P.,  et~al., 2019, The Astrophysical Journal, 872, 94

\bibitem[\protect\citeauthoryear{{Mathis}, {Rumpl}  \& {Nordsieck}}{{Mathis}
  et~al.}{1977}]{mathis1977a}
{Mathis} J.~S.,  {Rumpl} W.,   {Nordsieck} K.~H.,  1977, The Astrophysical
  Journal, 217, 425

\bibitem[\protect\citeauthoryear{{Mazzalay}, {Rodriguez-Ardila}  \&
  {Komossa}}{{Mazzalay} et~al.}{2010}]{mazzalay2010a}
{Mazzalay} X.,  {Rodriguez-Ardila} A.,   {Komossa} S.,  2010, Monthly Notices
  of the Royal Astronomical Society, 405, 1315

\bibitem[\protect\citeauthoryear{{Mehdipour} \& {Costantini}}{{Mehdipour} \&
  {Costantini}}{2018}]{mehdipour2018a}
{Mehdipour} M.,  {Costantini} E.,  2018, Astronomy \& Astrophysics, 619, 12

\bibitem[\protect\citeauthoryear{{Meléndez}, {Kraemer}, {Weaver}  \&
  {Mushotzky}}{{Meléndez} et~al.}{2011}]{melendez2011a}
{Meléndez} M.,  {Kraemer} S.~B.,  {Weaver} K.~A.,   {Mushotzky} R.~F.,  2011,
  The Astrophysical Journal, 738, 6

\bibitem[\protect\citeauthoryear{{Nagao}, {Taniguchi}  \& {Murayama}}{{Nagao}
  et~al.}{2000}]{nagao2000a}
{Nagao} T.,  {Taniguchi} Y.,   {Murayama} T.,  2000, The Astronomical Journal,
  119, 2605

\bibitem[\protect\citeauthoryear{{Netzer}}{{Netzer}}{2004}]{netzer2004a}
{Netzer} H.,  2004, \apj, 604, 551

\bibitem[\protect\citeauthoryear{{Ogle}, {Marshall}, {Lee}  \&
  {Canizares}}{{Ogle} et~al.}{2000}]{ogle2000a}
{Ogle} P.~M.,  {Marshall} H.~L.,  {Lee} J.~C.,   {Canizares} C.~R.,  2000, The
  Astrophysical Journal, 545, L81

\bibitem[\protect\citeauthoryear{{Porquet}, {Dumont}, {Collin}  \&
  {Mouchet}}{{Porquet} et~al.}{1999}]{porquet1999a}
{Porquet} D.,  {Dumont} A.-M.,  {Collin} S.,   {Mouchet} M.,  1999, Astronomy
  and Astrophysics, 341, 58

\bibitem[\protect\citeauthoryear{{Revalski} et~al.,}{{Revalski}
  et~al.}{2021}]{revalski2021a}
{Revalski} M.,  et~al., 2021, \mn@doi [The Astrophysical Journal]
  {10.3847/1538-4357/abdcad}, 910, 139

\bibitem[\protect\citeauthoryear{{Rodríguez-Ardila}, {Viegas}, {Pastoriza}  \&
  {Prato}}{{Rodríguez-Ardila} et~al.}{2002}]{rodriguez-ardila2002a}
{Rodríguez-Ardila} A.,  {Viegas} S.~M.,  {Pastoriza} M.~G.,   {Prato} L.,
  2002, The Astrophysical Journal, 579, 214

\bibitem[\protect\citeauthoryear{{Rodríguez-Ardila}, {Prieto}, {Portilla}  \&
  {Tejeiro}}{{Rodríguez-Ardila} et~al.}{2011}]{rodriguez-ardila2011a}
{Rodríguez-Ardila} A.,  {Prieto} M.~A.,  {Portilla} J.~G.,   {Tejeiro} J.~M.,
  2011, The Astrophysical Journal, 743, 100

\bibitem[\protect\citeauthoryear{{Romano} et~al.,}{{Romano}
  et~al.}{2004}]{romano2004a}
{Romano} P.,  et~al., 2004, The Astrophysical Journal, 602, 635

\bibitem[\protect\citeauthoryear{{Satyapal}, {Kamal}, {Cann}, {Secrest}  \&
  {Abel}}{{Satyapal} et~al.}{2021}]{satyapal2021a}
{Satyapal} S.,  {Kamal} L.,  {Cann} J.~M.,  {Secrest} N.,   {Abel} N.~P.,
  2021, The Astrophysical Journal, 906, 35

\bibitem[\protect\citeauthoryear{{Seyfert}}{{Seyfert}}{1943}]{seyfert1943a}
{Seyfert} C.~K.,  1943, The Astrophysical Journal, 97, 28

\bibitem[\protect\citeauthoryear{{Snow} \& {Witt}}{{Snow} \&
  {Witt}}{1996}]{snow1996a}
{Snow} T.~P.,  {Witt} A.~N.,  1996, The Astrophysical Journal, 468, L65

\bibitem[\protect\citeauthoryear{{Trindade Falcão} et~al.,}{{Trindade Falcão}
  et~al.}{2021a}]{trindadefalcao2021a}
{Trindade Falcão} A.,  et~al., 2021a, Monthly Notices of the Royal
  Astronomical Society, 500

\bibitem[\protect\citeauthoryear{{Trindade Falcão} et~al.,}{{Trindade Falcão}
  et~al.}{2021b}]{trindadefalcao2021b}
{Trindade Falcão} A.,  et~al., 2021b, Monthly Notices of the Royal
  Astronomical Society, 505

\bibitem[\protect\citeauthoryear{{Wang} et~al.,}{{Wang}
  et~al.}{2011a}]{wang2011b}
{Wang} J.,  et~al., 2011a, The Astrophysics Journal, 729, 75

\bibitem[\protect\citeauthoryear{{Wang}, {Fabbiano}, {Elvis}, {Risaliti},
  {Mundell}, {Karovska}  \& {Zezas}}{{Wang} et~al.}{2011b}]{wang2011a}
{Wang} J.,  {Fabbiano} G.,  {Elvis} M.,  {Risaliti} G.,  {Mundell} C.~G.,
  {Karovska} M.,   {Zezas} A.,  2011b, The Astrophysics Journal, 736, 62

\bibitem[\protect\citeauthoryear{{Wang} et~al.,}{{Wang}
  et~al.}{2011c}]{wang2011c}
{Wang} J.,  et~al., 2011c, The Astrophysics Journal, 742, 23

\bibitem[\protect\citeauthoryear{{Young}, {Wilson}  \& {Shopbell}}{{Young}
  et~al.}{2001}]{young2001a}
{Young} A.~J.,  {Wilson} A.~S.,   {Shopbell} P.~L.,  2001, The Astronomical
  Journal, 556, 6

\bibitem[\protect\citeauthoryear{{Zhang} \& {Sampson}}{{Zhang} \&
  {Sampson}}{2002}]{zhang2002a}
{Zhang} H.~L.,  {Sampson} D.~H.,  2002, \mn@doi [Atomic Data and Nuclear Data
  Tables] {https://doi.org/10.1006/adnd.2002.0888}, 82, 357

\makeatother
\end{thebibliography}


\appendix

\section{Appendix A}
\label{sec:appendix}

\begin{figure*}
  \centering
 \begin{minipage}[b]{0.45\textwidth}
 \hspace{-3mm}
  \includegraphics[width=8.65cm]{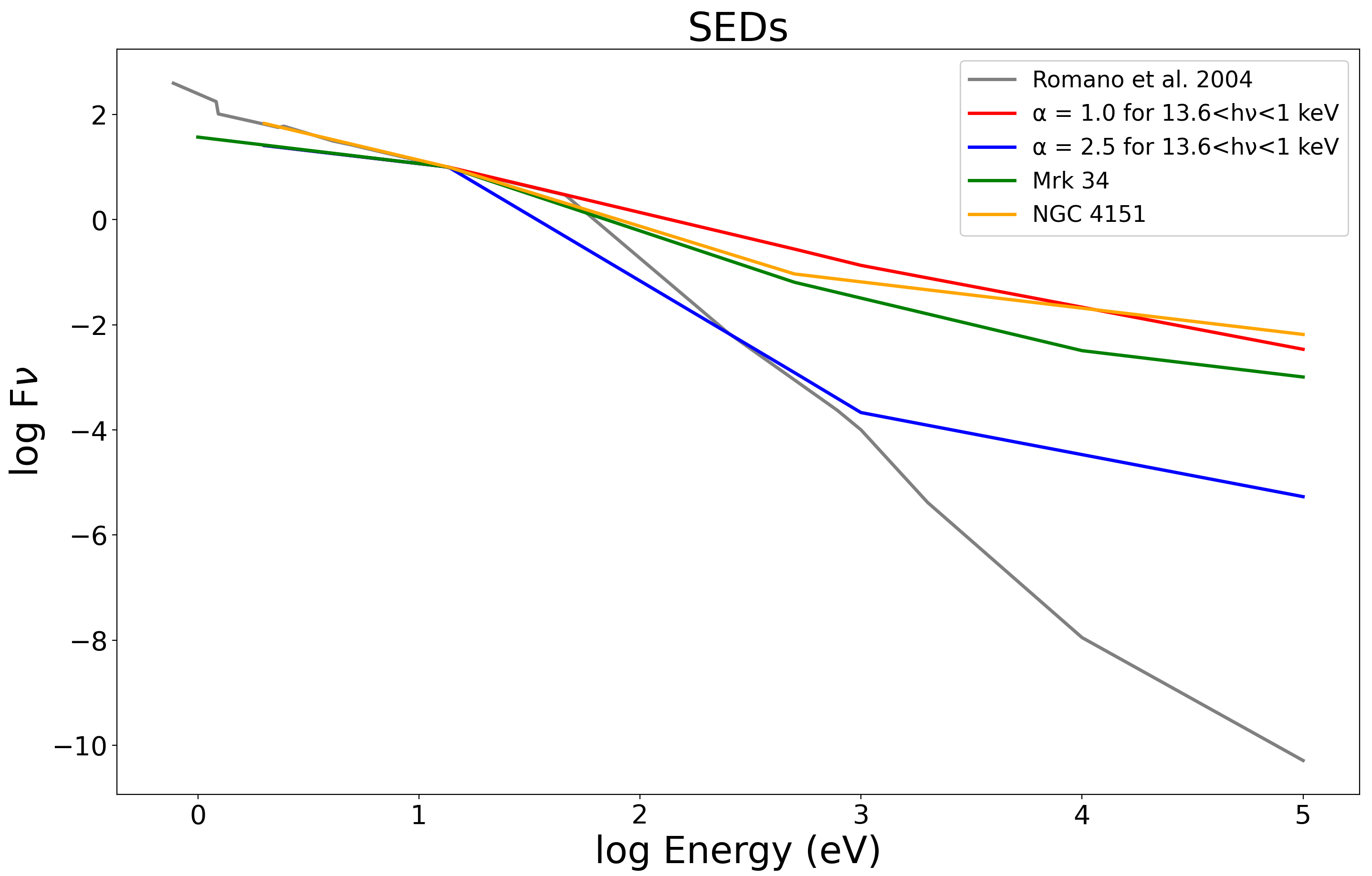}
 \end{minipage}\qquad 
 \begin{minipage}[b]{0.45\textwidth}
 \hspace{-3mm}
  \includegraphics[width=8.65cm]{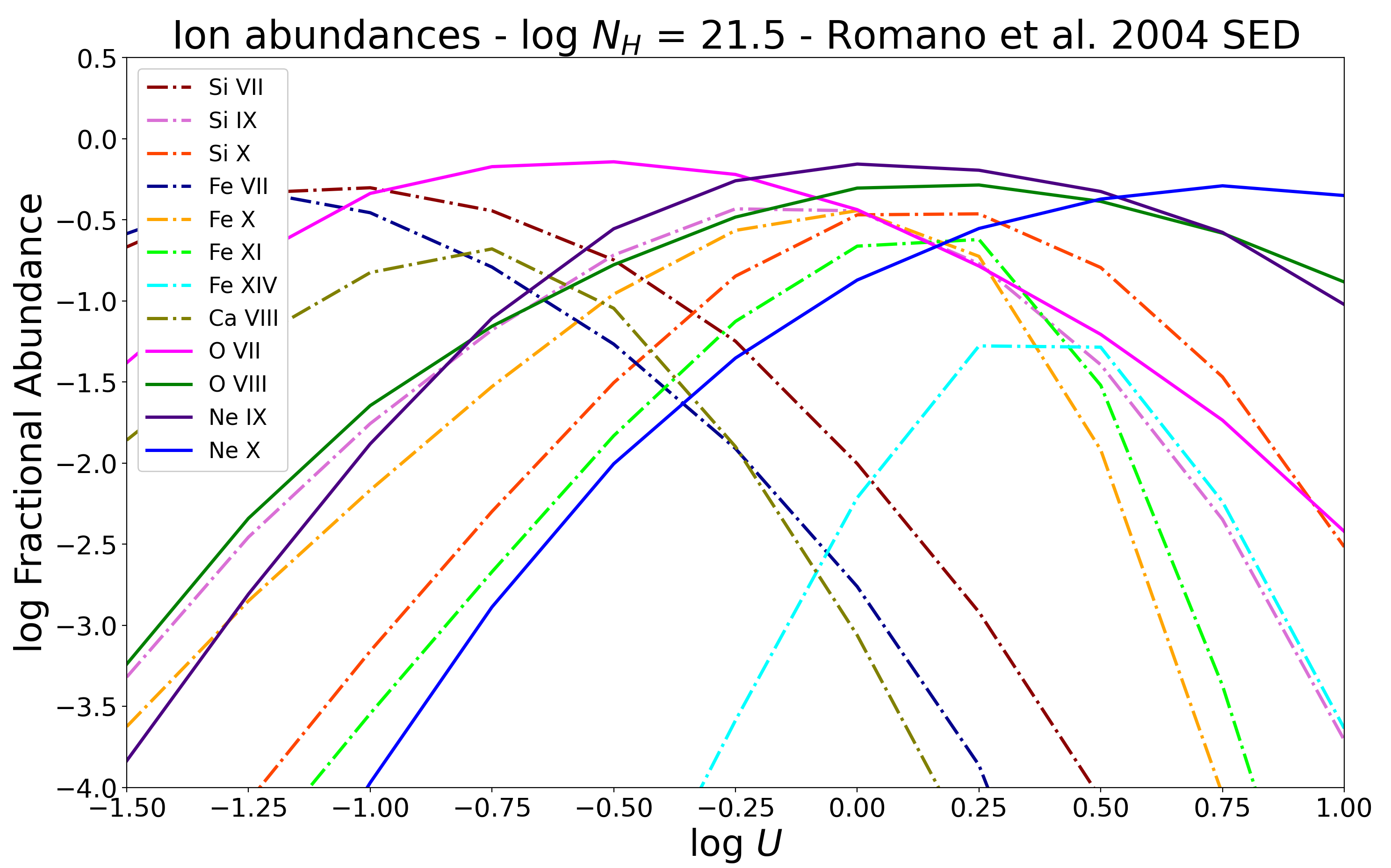}
 \end{minipage}\qquad 
 \begin{minipage}[b]{0.45\textwidth}
  \includegraphics[width=8.65cm]{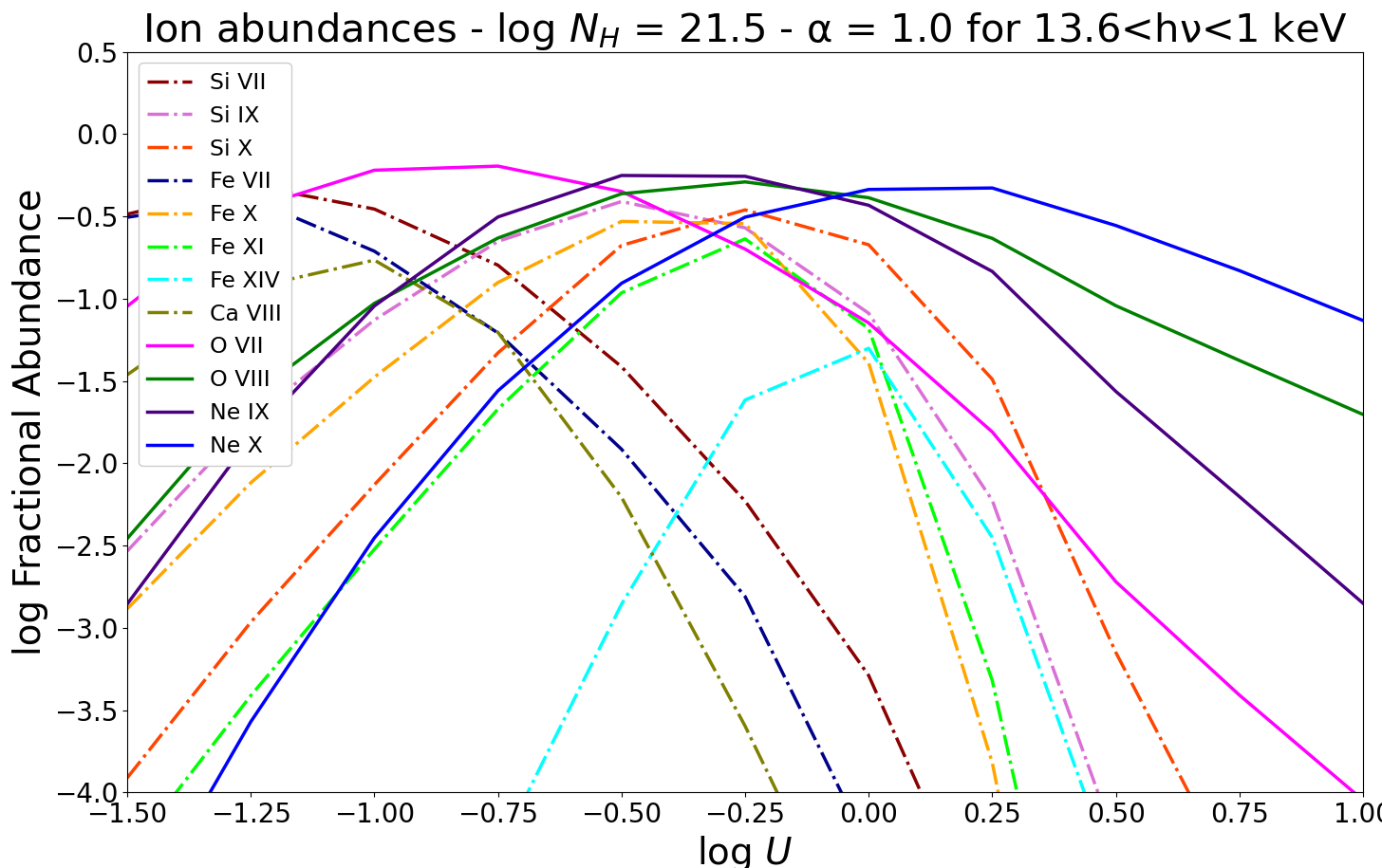}
\end{minipage}\qquad 
 \begin{minipage}[b]{0.45\textwidth}
  \includegraphics[width=8.65cm]{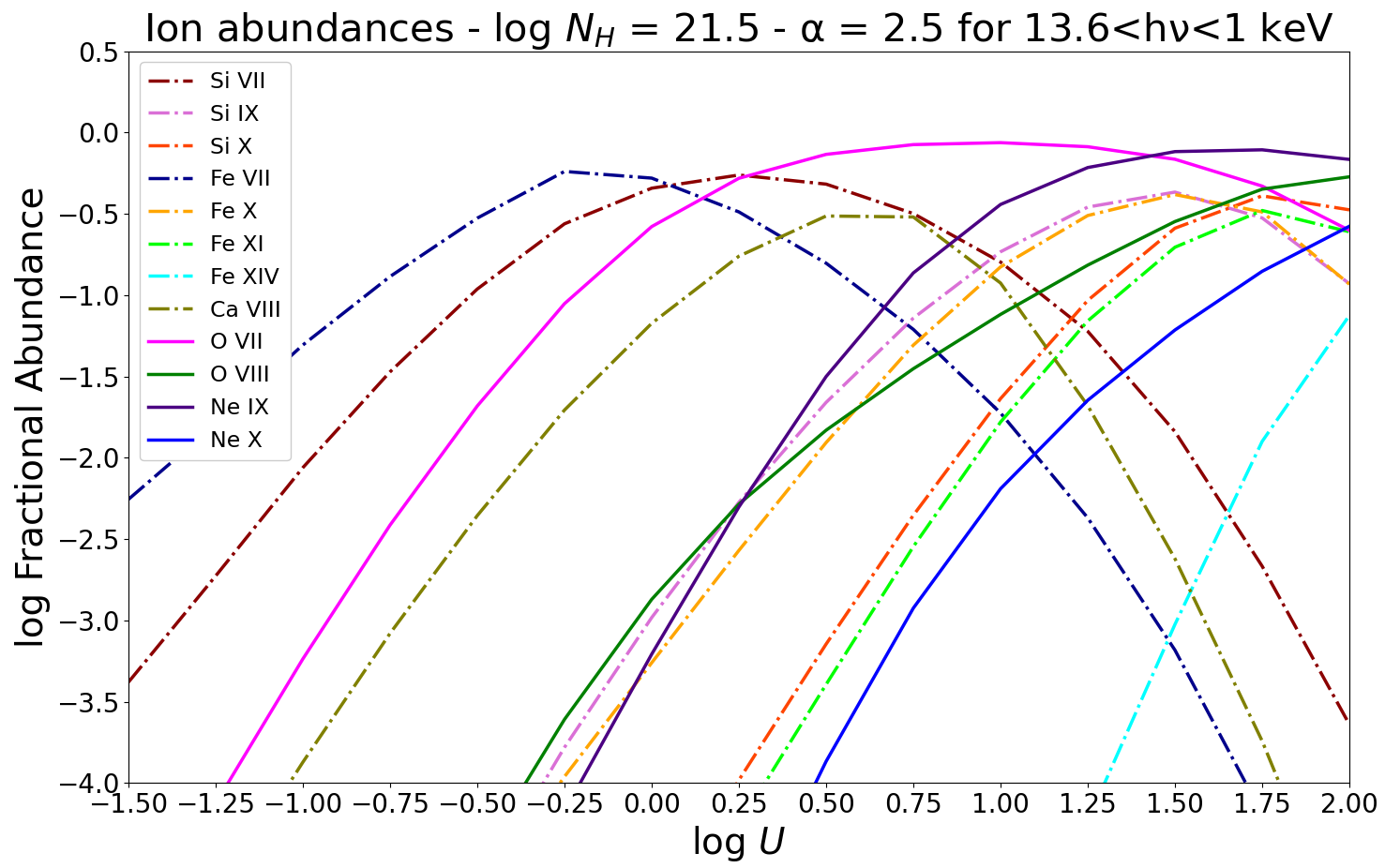}
 \end{minipage}\qquad 
\caption{\textbf{\textit{Top left figure:}} Representation of the five SEDs used in our analysis. \textbf{\textit{Top right figure:}} Cloudy model predictions for the fractional abundances of relevant ionisation states of iron and silicon and H- and He-like oxygen and neon. The model assumptions are: abundances (1.4x solar) and spectral energy distribution (SED) from \citealt{romano2004a}, and constant column density N$_{H}$ = 10$^{21.5}$ cm$^{-2}$ \citep[e.g.,][]{kraemer2020a, trindadefalcao2021b}. As shown here, optical emission lines from ions such as Fe~X, Fe~XI, Fe~XIV, and IR emission lines from ions such as Si~X will be formed in the X-ray emitting gas and will act as X-ray ``footprints" \textbf{\textit{Bottom left figure:}} Same as top right figure, but with an SED from \citealt{melendez2011a} with $\alpha$ = 1.0 between 13.6 $<~ h\nu~<$~1 keV. \textbf{\textit{Bottom right figure:}} Same as top right figure, but with an SED from \citealt{melendez2011a} with $\alpha$ = 2.5 between 13.6 $<~ h\nu~<$~1 keV. The results shown in the top-right, bottom-left, and bottom-right figures can be compared to our results using the Mrk 34 SED (Figure \ref{fig:cloudy}) and the results from the NGC 4151 analysis (see Table \ref{tab:fluxes}).}
\label{fig:SED}
\end{figure*}

As noted in Section \ref{sec:study_footprints}, we also explore the effect of different SEDs on the fractional abundances of relevant ionisation states of iron and silicon and H- and He- like oxygen and neon. To perform this analysis we use the SED present by \citet{romano2004a} and the range of power-law SEDs presented by \citet{melendez2011a}. The different SEDs used and our results are shown in Figure \ref{fig:SED}.\par 

In Figure \ref{fig:SED} we show the results of our analysis for 
$\alpha$ = 1.0 between 13.6 $<h\nu<$ 1 keV, and $\alpha$ = 2.5 between 13.6 $<h\nu<$ 1 keV, i.e., the lower and upper values of the range modeled by \citet{melendez2011a}\footnote{The SEDs modeled by \citet{melendez2011a} are characterised by $\alpha$ = 0.5 for $h\nu <$ 13.6 eV, $\alpha$ = 1.0 - 2.5 between 13.6 $<~ h\nu~<$~1 keV, and $\alpha$ = 0.8 for $h\nu>$1 keV}. As we can see in Figure \ref{fig:SED}, the ions from which the footprint lines originate and H- and He- like ions still coexist over the same range in ionisation parameter, even when considering sources with various SEDs. Note that for $\alpha$ = 2.5 between 13.6 $<h\nu<$ 1 keV, there are relatively fewer EUV and X-ray photons, therefore the fractional abundances peak at higher ionisation parameter than for the harder SEDs, as shown at the bottom right panel in Figure \ref{fig:SED}. However, it is possible to see that the optical/IR footprint lines still coexist over the same range in ionisation parameter as the X-ray lines, which confirms the accuracy of our method, even for very soft SEDs. 

\bsp	
\label{lastpage}
\end{document}